\documentclass[superscriptaddress,aps,pra,twocolumn]{revtex4}
\usepackage{bm}
\usepackage{ulem}
\usepackage{epsfig}
\usepackage{graphicx}
\usepackage{amssymb,amsmath,amsbsy,amsgen,amsfonts}    
\usepackage{dcolumn}
\usepackage{amsthm}
\usepackage{mathrsfs}
\usepackage{latexsym}
\usepackage{array}
\usepackage{color}
\usepackage{amstext}
\allowdisplaybreaks[1]
\usepackage{txfonts}
\usepackage{pifont}
\usepackage{braket}
\usepackage{epstopdf} 

\newcommand{\be}{\begin{equation}}
\newcommand{\ee}{\end{equation}}
\newcommand{\ba}{\begin{array}}
\newcommand{\ea}{\end{array}}
\newcommand{\bqa}{\begin{eqnarray}}
\newcommand{\eqa}{\end{eqnarray}}

\DeclareSymbolFont{symbols}{OMS}{cmsy}{m}{n}

\begin{document}

\title{Measuring kinetic parameters using quantum plasmonic sensing}

\author{K. T. Mpofu}
\affiliation{Catalysis\,and\,Peptide\,Research\,Unit,\,School\,of\,Health\,Sciences,\,University\,of\,KwaZulu-Natal,\,Durban\,4041,\,South\,Africa}

\author{C. Lee}
\affiliation{Quantum Universe Center, Korea Institute for Advanced Study, Seoul 02455, Republic of Korea}
\affiliation{Korea Research Institute of Standards and Science, Daejeon 34113, Republic of Korea}

\author{G. E. M. Maguire}
\affiliation{Catalysis\,and\,Peptide\,Research\,Unit,\,School\,of\,Health\,Sciences,\,University\,of\,KwaZulu-Natal,\,Durban\,4041,\,South\,Africa}
\affiliation{School of Chemistry and Physics, University of KwaZulu-Natal, Durban 4041, South Africa}

\author{H. G. Kruger}
\affiliation{Catalysis\,and\,Peptide\,Research\,Unit,\,School\,of\,Health\,Sciences,\,University\,of\,KwaZulu-Natal,\,Durban\,4041,\,South\,Africa}

\author{M. S. Tame}
\affiliation{Laser Research Institute, Department of Physics, Stellenbosch University, Private Bag X1, Matieland 7602, South Africa}

\date{\today}

\begin{abstract}
The measurement of parameters that describe kinetic processes is important in the study of molecular interactions. It enables a deeper understanding of the physical mechanisms underlying how different biological entities interact with each other, such as viruses with cells, vaccines with antibodies, or new drugs with specific diseases. In this work, we study theoretically the use of quantum sensing techniques for measuring the kinetic parameters of molecular interactions. The sensor we consider is a plasmonic resonance sensor -- a label-free photonic sensor that is one of the most widely used in research and industry. The first type of interaction we study is the antigen BSA interacting with antibody IgG1, which provides a large sensor response. The second type is the enzyme carbonic anhydrase interacting with the tumor growth inhibitor benzenesulfonamide, which produces a small sensor response. For both types of interaction we consider the use of two-mode Fock states, squeezed vacuum states and squeezed displaced states. We find that these quantum states offer an enhancement in the measurement precision of kinetic parameters when compared to that obtained with classical light. The results may help in the design of more precise quantum-based sensors for studying kinetics in the life sciences.
\end{abstract}

\maketitle

\section{Introduction}

Measuring parameters used in kinetic models of molecular interactions provides important information about how pathogens, such as viruses and bacteria, interact with cells and other biological entities~\cite{Pollard2010}. Plasmonic sensors have long been used to measure kinetic parameters in this context due to their higher sensitivity compared to conventional sensors~\cite{Homola1999,Homola1999b,Homola2003,Homola2006,Li2015,Salazar2018,Xiao19}. Another main benefit of using plasmonic sensors is that the gold surface used in a given sensor allows for a straightforward immobilization of different types of molecules by appropriate functionalization~\cite{Homola2008}. This enables a controlled setting in which to investigate the binding and unbinding (release) of a specific type of molecule, or biological entity, to another in a label-free manner~\cite{Soler2019}. Despite the successful performance of plasmonic sensors for measuring kinetics in a wide range of scenarios~\cite{Xiao19}, the precision in the estimation of kinetic parameters is reaching a fundamental classical limit, given by the shot noise of the classical light source used in the sensors~\cite{Piliarik2009,Wang2011}. This is particularly the case for binding and unbinding processes that elicit a small response in the sensor's signal, for instance between a virus and an inhibitor drug~\cite{Shankaran2007}. 

Fortunately, quantum techniques have been developed recently for plasmonic sensors using quantum light sources and measurements that enable a significant reduction of noise below the shot-noise limit leading to a better precision in the estimation of a parameter being sensed~\cite{Lee2021}. So far, these quantum techniques have been applied to the measurement of static parameters, such as the concentration of a substance via a measurement of the induced change in the refractive index surrounding a plasmonic sensor. Theoretical studies have looked at the use of different types of quantum states of light and measurements for the precise estimation of the refractive index change~\cite{Lee2017,Tame19}. Related experiments have used quantum states to demonstrate an enhancement in the precision of the estimation of bovine serum albumin (BSA) concentration; using single photons~\cite{Lee2018,Peng2020,Zhao2020} and squeezed states~\cite{Fan2015,Pooser2016,Dowran2018}. However, it is currently unclear as to whether the enhancement in the precision of estimating a static parameter, such as the concentration, translates to a parameter that defines the dynamics of a physical process.

In this work we study theoretically the measurement of the interaction kinetics of two physical processes using a plasmonic resonance sensor and quantum states of light. The first process we study is the antigen BSA interacting with the antibody rabbit anti-cow albumin IgG1 (anti-BSA). This interaction is well documented in the work of Kausaite {\it et al.}~\cite{Kausaite07}. It was chosen as it provides a large sensor response and thus acts as a basic starting point for the study of quantum techniques for measuring kinetic parameters. The second process we study is carbonic anhydrase interacting with benzenesulfonamide, which is well studied in the work of Lahiri {\it et al.}~\cite{Lahiri99}. Carbonic anhydrase is an enzyme used for maintaining pH balance in blood and it also facilitates the removal/transport of carbon dioxide from cells via red blood cells.  Benzenesulfonamide can be used as an inhibitor of the enzyme and acts as an efficient tumor growth inhibitor~\cite{Supuran2000}. This second process was chosen as it produces a small sensor response and therefore gives information about the performance of quantum techniques for measuring kinetic parameters in pathogen-inhibitor interactions relevant to drug development~\cite{Shankaran2007,Homola2008}. 
\begin{figure*}[t]
\centering
\includegraphics[width=17.5cm]{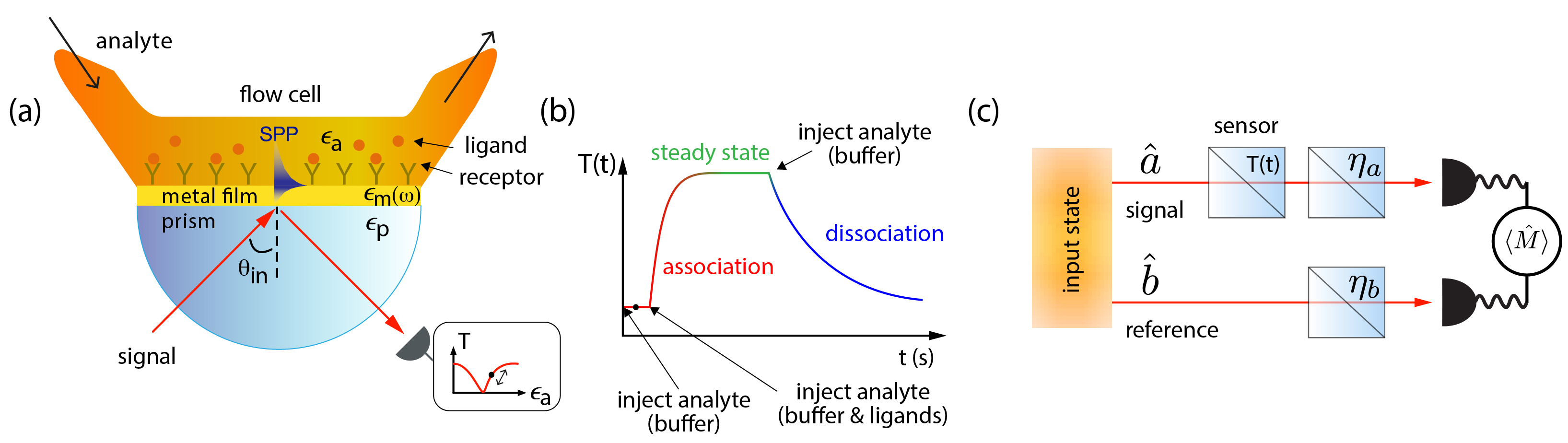}
\caption{Measuring interaction kinetics using a quantum plasmonic sensor. (a) Surface plasmon resonance sensor with flow cell for bringing an analyte into and out of the sensing region above a gold surface (metal film). Inset shows how the transmittance, $T$, of the sensor changes as the refractive index $n_a$ (or permittivity $\epsilon_a=n_a^2$) above the gold surface varies due to the presence of the analyte and a surface plasmon polariton (SPP). (b) Sensor signal, known as a sensorgram, during the kinetics as the flow cell is injected with the analyte. The first phase is association as the ligands in the analyte bind to the receptors immobilized on the gold surface. Mass transport effects due to diffusion are neglected in the model. The second phase is steady state when the binding and unbinding rates are equal. The final phase is dissociation as a buffer solution is injected into the flow cell to remove the ligands. (c) The surface plasmon resonance sensor can be placed in a two-mode setup in order to exploit quantum sensing techniques. Here, it plays the role of a beamsplitter with time-varying transmittance, $T(t)$. Loss on mode a (b) due to experimental imperfections can be modelled as an additional beamsplitter with transmittance $\eta_a$ ($\eta_b$).}
\label{fig1} 
\end{figure*}

The kinetic parameters in both processes studied are obtained by measuring the intensity in one mode (which includes the sensor) or intensity difference of two modes (one with the sensor and the other as a reference) over time. The corresponding temporal measurement signal forms what is called a sensorgram and allows the kinetic parameters to be extracted via a nonlinear fit. A key requirement of this dynamical sensing method is that the sensorgram varies on a timescale much longer than the timescale of probing with quantum states of light. This allows many probe states to be sent into the sensor to build up the required statistics at each point in time. In both processes studied, due to the kinetic parameter values, the sensorgram changes slowly, on the order of seconds, while the assumption is that, for example, 1000 probe quantum states can be prepared and sent into the sensor per second. This is a reasonable assumption given current experimental capabilities~\cite{Lee2018,Peng2020,Zhao2020,Fan2015,Pooser2016,Dowran2018}.

For both types of interaction studied we consider various quantum states: two-mode Fock states, two-mode squeezed vacuum states and two-mode squeezed displaced states, all used in a recent study of static parameters~\cite{Lee2017,Tame19}. Our results show that the enhancement in the estimation precision offered by these quantum states translates well from static to kinetic parameters under certain conditions. In particular, we compare the estimation precision of the association and dissociation binding constants of the interactions using classical and quantum states. 

It is important to point out that in the classical case that we use as a benchmark in our study, {\it i.e.}, a two-mode coherent state, the estimation precision is set by the shot noise, which is inversely proportional to the square root of the intensity~\cite{Lee2021}. Therefore one could simply increase the intensity per probe state, {\it i.e.}, mean photon number in the modes, and thereby decrease the noise to obtain a better precision. The key point here is that for a fixed mean photon number per state and fixed number of states per second, it is known in the static case that there are quantum states that always outperform the two-mode classical coherent state in terms of providing a better precision. Thus, one is able to obtain the same estimation precision as the classical state using a quantum state with a reduced intensity. This is of the utmost importance when the biological sample is photosensitive~\cite{Casacio2021}, or the sensor is operating near its intensity limit in providing a linear response~\cite{Piliarik2009,Kaya2013}. It is in this setting where a static quantum plasmonic sensor would provide a practical advantage and in our work we show that this is also the case for a dynamic quantum plasmonic sensor. Our work may therefore potentially aid the future design of plasmonic sensors using quantum techniques for more precise kinetic research where the above factors play a role.

The work is organized as follows: In Section II, we introduce the physical model we consider for plasmonic sensing, where we provide details of the sensor setup, its signal response to a dynamically changing environment, as well as the various quantum states and measurements we study. In Section III, we then discuss the general model for interaction kinetics used and show how the kinetic parameters can be measured, for a given interaction, from the plasmonic sensor's signal. We also outline the simulation method we use to model noise in the signal for the different quantum states and measurements, thereby simulating a potential experiment. In Sections IV and V, we present our results for the two molecular interaction processes studied. In Section IV a large response in the sensor signal is studied and in Section V a small response is studied. In Section VI we summarize our findings.

\section{Sensing model}

\subsection{SPR setup and resonance dip}

In Fig.~\ref{fig1}(a) we show the setup we consider for measuring interaction kinetics, known as the Kretschmann configuration~\cite{Kret1968,Kret1971}. Here, light in a signal mode is incident on a prism and interacts with a thin metal layer (metal film) on top of the prism. With the correct coupling conditions the light in the signal mode excites conduction electrons on the upper surface of the metal. This creates a surface electromagnetic wave -- a surface plasmon polariton (SPP) -- that is confined to the upper metal surface. The result of this SPP excitation is represented as a drop in the intensity of the reflected light in the signal mode, which is a phenomenon known as surface plasmon resonance (SPR). For the setup to act as a sensor based on SPR, {\it i.e.}, a plasmonic sensor, a metal with free electrons in the conduction band is needed. In our case, we have chosen gold as it is regularly used in plasmonic experiments due to its overall stability~\cite{Homola2006}. 

A dielectric material whose refractive index, $n_a=\sqrt{\epsilon_a}$ (where $\epsilon_a$ is the permittivity), is to be sensed is placed on top of the gold. The `material' in our case is a mix of receptors immobilized on the gold surface and ligands that bind to them introduced in an analyte via a flow cell. The optimal coupling of light to an SPP on the gold surface occurs for a fixed value of the receptor-ligand permittivity, $\epsilon_a$, together with a polarization of light in the plane of incidence and a specific angle of incidence, $\theta_{in}$. Under these conditions the reflected light in the signal mode is reduced. The reflected light can also be thought of as the light that is `transmitted' through the sensor by considering the sensor as a device with one input and one output. Thus, for optimal coupling the transmittance, $T$, of the signal mode through the plasmonic sensor is reduced. In the ideal case, the transmittance reduces to zero, corresponding to a complete conversion of light to an SPP. When $\epsilon_a$ varies, as a result of changes in how many ligands are binding or unbinding to receptors on the surface, the transmittance changes, as shown in the inset of Fig.~\ref{fig1}(a). Thus, the receptor-ligand interaction process can be monitored indirectly by measuring the transmitted light in the signal mode with a photo-detector that measures the intensity. 

In order to ensure the highest sensitivity to changes in $\epsilon_a$, the incident angle $\theta_{in}$ is set such that for an analyte with no ligands (known as the buffer solution) the transmittance $T$ is at the mid-point of the resonance curve (inflection point), shown as a black dot on the right hand side of the inset of Fig.~\ref{fig1}(a). The analyte flows over the metal surface at a constant rate and $\epsilon_a$ remains constant, leading to a constant $T$, as shown in the first stage of Fig.~\ref{fig1}(b). When ligands are added to the injected analyte the value of $\epsilon_a$ increases due to binding of ligands to the receptors as the analyte flows over the surface. This is a dynamic process where the ligands are continuously adsorbed/associated/immobilised onto the receptors and again desorbed/dissociated from them. The transmittance $T$ therefore increases (see inset of Fig.~\ref{fig1}(a)) and gains a time dependence. This is shown as the association stage for $T$ in Fig.~\ref{fig1}(b). During this stage, ligands and receptors bind and unbind (are released). A steady state is eventually reached as the analyte with ligands continues to flow at a constant rate. Finally, pure analyte is then `pumped' through the cell, where the ligands are removed from the analyte so that only the buffer solution (pure analyte) flows across the surface. The ligands bound to the receptors gradually unbind and are `washed' away from the surface so that it is unlikely any further binding takes place, causing $\epsilon_a$ to reduce and $T$ to decrease. This is shown as the dissociation stage for $T$ in Fig.~\ref{fig1}(b). The complete curve for $T(t)$ as a function of time is known as a sensorgram and from it the kinetic parameters can be extracted. Further details about the kinetics and the method of parameter extraction will be given in the next section. Here, we briefly provide details on the general operation of the sensor in terms of a changing $\epsilon_a$.

The above type of plasmonic sensing is called intensity interrogation and it is complementary to another type of sensing called angular interrogation~\cite{Homola1999,Homola1999b}, which is also widely used, where for a given $\epsilon_a$ the input angle $\theta_{in}$ is varied and the angle causing a minimum in the transmittance $T$ is used to infer the value of $\epsilon_a$. Both types of interrogation offer a similar sensitivity and performance~\cite{Lee2021}. We have chosen to focus on intensity interrogation as it is more amenable for incorporating quantum sensing techniques and analyzing their performance. 

Formally, the transmittance of the light in the signal mode can be found using a 3-layer model for the sensor and is expressed mathematically as $T=|r_{spp}|^2$, where $r_{spp}$ is the reflection coefficient given by~\cite{Raether1986}
\begin{equation}
r_{spp}=\frac{e^{i2k_{2}d} r_{23} + r_{12}}{e^{i2k_{2}d} r_{23} r_{12} + 1},
\label{eqn:r}
\end{equation}
where 
\begin{equation}
r_{uv}=\left(\frac{{k_{u}}}{\epsilon_{{u}}} -\frac{{k_{v}}}{\epsilon_{{v}}}\right) \bigg/\left(\frac{{k_{u}}}{\epsilon_{{u}}} + \frac{{k_{v}}}{\epsilon_{{v}}}\right).
\end{equation}
In the above, $d$ is the thickness of the metal film, $k_i=\sqrt{\epsilon_i} (\omega/c)[1-(\epsilon_1/\epsilon_i)\sin^2\theta_{in}]^{1/2}$ is the normal-to-surface component of the wavevector in the $i$-th layer, $\omega$ is the angular frequency of the light in the signal mode and $\epsilon_i$ is the respective permittivity, where $\epsilon_1=\epsilon_p$, $\epsilon_2=\epsilon_m$ and $\epsilon_3=\epsilon_a$ for the layers.
 
\subsection{Quantum states considered}

Typically in the Kretschmann configuration, as described above, the source of light is a laser. This source of classical light is well represented as a coherent state in the quantum formalism~\cite{Loudon2000,Wiseman2016}. Thus, when comparing the different sources of light for measuring the kinetics we use the coherent state as the classical benchmark. This leads to the classical shot-noise limit (SNL), which is a fundamental limit for the precision of the sensor due to noise stemming from the statistical nature of the laser light~\cite{Loudon2000,Lee2021}. In Fig.~\ref{fig1}(a) the plasmonic sensor uses a single mode for sensing, however in recent years the use of a second mode as a reference has been considered for removing common excess noise in order to reach the fundamental SNL for classical light~\cite{Wu2004,Piliarik2009,Wang2011}. This more general two mode setting is shown in Fig.~\ref{fig1}(c) and modelled using beamsplitters, with a signal mode $a$ and a reference mode $b$. Here, the plasmonic sensor is represented by a beamsplitter with transmittance coefficient $T$. For realistic modelling, losses in mode $a$ ($b$) are represented by an additional beamsplitter, with transmittance coefficient $\eta_a$ ($\eta_b$). Finally, a differential measurement is made in order to estimate the transmittance $T$ set by the sensor's response. More details about this measurement are given in the next subsection.

The input state in the classical case is a coherent state of light in each mode, a two-mode coherent (TMC) state, which is represented as
\begin{equation}
\ket{\rm TMC}=\ket{\alpha}_a \ket{\beta}_b = {\hat{D}_a}(\alpha) {\hat{D}_b}(\beta) \ket{0}_a \ket{0}_b, \label{TMC}
\end{equation}
where $\hat{{D}}_a(\alpha) = {e}^{\alpha \hat{{a}}^{\dagger} - {\alpha}^* \hat{{a}}}$ is the displacement operator for mode $a$, with displacement parameter $\alpha \in {\mathbb C}$, and $\hat{a}^\dag$ and $\hat{a}$ as creation and annihilation operators for mode $a$, which obey the bosonic commutation relation $[\hat{a},\hat{a}^\dag]=1$~\cite{Loudon2000}. A similar expression can be written for the displacement operator of mode $b$. The displacement in Eq.~\eqref{TMC} creates a coherent state $\ket{\alpha}$ in mode $a$ with mean photon number $\langle \hat{N}_a \rangle=\langle \hat{a}^\dag\hat{a} \rangle=|\alpha|^2$ and similarly a coherent state $\ket{\beta}$ in mode $b$ with $\langle \hat{N}_b \rangle=|\beta|^2$. In what follows, in order to put the different quantum states on an equal footing with the classical TMC state we set $|\alpha|^2$ and $|\beta|^2$ to match the mean photon number in the signal and reference modes. In the case of states with an equal number of photons in either mode (balanced case), we have simply $|\alpha|^2=|\beta|^2=N$, where $N$ is the mean number of photons in either input mode.

In this two mode setting, it is possible to take advantage of quantum states of light that have inter-mode correlations in order to reduce the measurement noise below the SNL. A recent study has considered various two-mode quantum states for plasmonic sensing of a static quantity~\cite{Lee2017,Tame19}. Here, we select the best performing of the quantum states from that work and go beyond it by studying their ability to reduce the measurement noise below the SNL of a dynamic quantity and thereby provide a more precise measurement of kinetic parameters.

The first quantum state we consider is the two-mode Fock (TMF) state, which is expressed as
\begin{equation}
\ket{\rm TMF}=\ket{{N}}_a \ket{{N}}_b = \frac{(\hat{{a}}^{\dagger})^N}{\sqrt{{N!}}} \frac{(\hat{{b}}^{\dagger})^N}{\sqrt{{N!}}} \ket{0}_a \ket{0}_b.
\end{equation}
The TMF state has $N$ photons in each mode, and thus a mean photon number of $\langle \hat{N}_{a} \rangle=N$ and $\langle \hat{N}_{b} \rangle=N$ in modes $a$ and $b$, respectively. The generation of $N$-photon Fock states has been studied experimentally using linear optics~\cite{Zapletal2020}, atoms in cavities~\cite{Varcoe2000,Bertet2002,Varcoe2004,Zhou2012}, artificial quantum emitters~\cite{Waks2006} and superconducting quantum circuits~\cite{Wang2008,Hofheinz2008}. Theoretical schemes for generation have also been proposed and include linear optics~\cite{Clausen2001,Sanaka2005}, atoms in cavities~\cite{Groiseau2020}, and artificial emitters~\cite{Fischer2018,Uria2020,Cosacchi2020}.

The second quantum state we consider is the two-mode squeezed vacuum (TMSV) state, which is expressed as
\begin{equation}
\ket{\mathrm{TMSV}}={\hat{S}_{ab}(\chi)}\ket{0}_a\ket{0}_b,
\end{equation}
where ${\hat{S}_{ab}(\chi)}= e^{{\chi}^{\ast} {\hat{a} \hat{b}-{{\chi}} \hat{a}^{\dagger}\hat{b}^{\dagger}}}$ is a squeezing operation applied to the vacuum state. The squeezing parameter $\chi=r e^{i \theta_s}$, where $r$ represents the amount of squeezing and $\theta_s$ is a phase. The mean photon number in the modes is given by $\langle \hat{N}_{a} \rangle=\langle \hat{N}_{b} \rangle=\sinh^2 r=N$, with the value of $N$ set by the squeezing parameter $r$. The TMSV state has been used in many experiments to date~\cite{Meda2017} and can be generated optically using spontaneous parametric downconversion~\cite{Burnham1970,Heidmann1987,Schumaker1985}. Although more experimentally accessible than the Fock state for a given mean photon number $N$, for practical reasons the number of photons in each of the two modes is usually limited to small values of $N$.

For reaching higher $N$, the third quantum state we consider is the two-mode squeezed displaced (TMSD) state, which is expressed as
\begin{equation}
\ket{\mathrm{TMSD}}={\hat{S}_{ab}(\chi)}\ket{\alpha}_a\ket{0}_b.
\end{equation}
The mean photon number in each mode is $\langle \hat{N}_{a} \rangle=\sinh^2 r + \vert \alpha \vert^2 \cosh^2 r$ and~$\langle \hat{N}_{b} \rangle= \sinh^2 r + \vert \alpha \vert^2 \sinh^2 r$, which are dependent on the squeezing parameter $r$ and the initial value of the coherent state mean photon number $|\alpha|^2$. We set $|\alpha|^2$ and $r$ such that $\langle \hat{N}_{a} \rangle=N$, which then gives $\langle \hat{N}_{b} \rangle=N-|\alpha|^2$. The TMSD state has been studied in several works~\cite{Marino2009,Cai2015,Fang2015} and can be generated via a four-wave mixing process~\cite{McCormick2008,Boyer2008,Turnbull2013}, where intensities of up to several tens of $\mu$W (effectively very high $N$) have been achieved at the expense of reducing the photon-number correlation between the two modes in experiments for quantum sensing~\cite{Pooser2015,Clark2012}. We use $\cosh^2 r=4.5$ as a practical value in our study for the squeezing~\cite{Pooser2016} and set $|\alpha|^2$ of the coherent state appropriately in order to satisfy $\langle \hat{N}_{a} \rangle=N$. Note that the energy constraint of $N$ photons in the signal mode is imposed on all the cases we consider in this work. This leads to a fair comparison among the uses of the different probe states. 

\subsection{Parameter estimation}

In order to estimate the kinetic parameters from the temporal signal $T(t)$ shown in Fig.~\ref{fig1}(b), {\it i.e.}, the sensorgram, the transmittance $T$ must be estimated at a given time. The sensitivity of the measurement of $T$ from a measurement $\langle \hat{M} \rangle$ of some observable $\hat{M}$ is given by 
\be
S_{\!M}=\left| \frac{d\langle\hat{M}\rangle}{dT}\right|.
\ee
It can be understood as the extent to which the measurement expectation value $\langle \hat{M} \rangle$ changes for a given change in the transmittance $T$. The precision in the estimation of $T$ for a single-shot measurement $\langle \hat{M} \rangle$ is then
\be
\Delta T=\Delta M/S_{\!M},
\label{DeltaT}
\ee
where $\Delta M=(\langle\hat{M}^2\rangle-\langle\hat{M}\rangle^2)^{1/2}$ is the uncertainty (precision) of the measurement. For a sample consisting of a set of $\nu$ measurements, which leads to an estimate $\bar{T}$ using the mean as an estimator, the estimation precision of $\bar{T}$ becomes $\Delta T/\sqrt{\nu}$, which simply expresses that the estimation of $\bar{T}$ becomes more precise as the sample size increases.

Theoretically, $\Delta M$ and $S_{\!M}$ depend on the quantum state used, the type of measurement and the sensor setup~\cite{Lee2017}. With these things all fixed, the estimation precision $\Delta T$ can then be calculated from Eq.~\eqref{DeltaT}.
\begin{figure}[t]
\centering
\includegraphics[width=8.8cm]{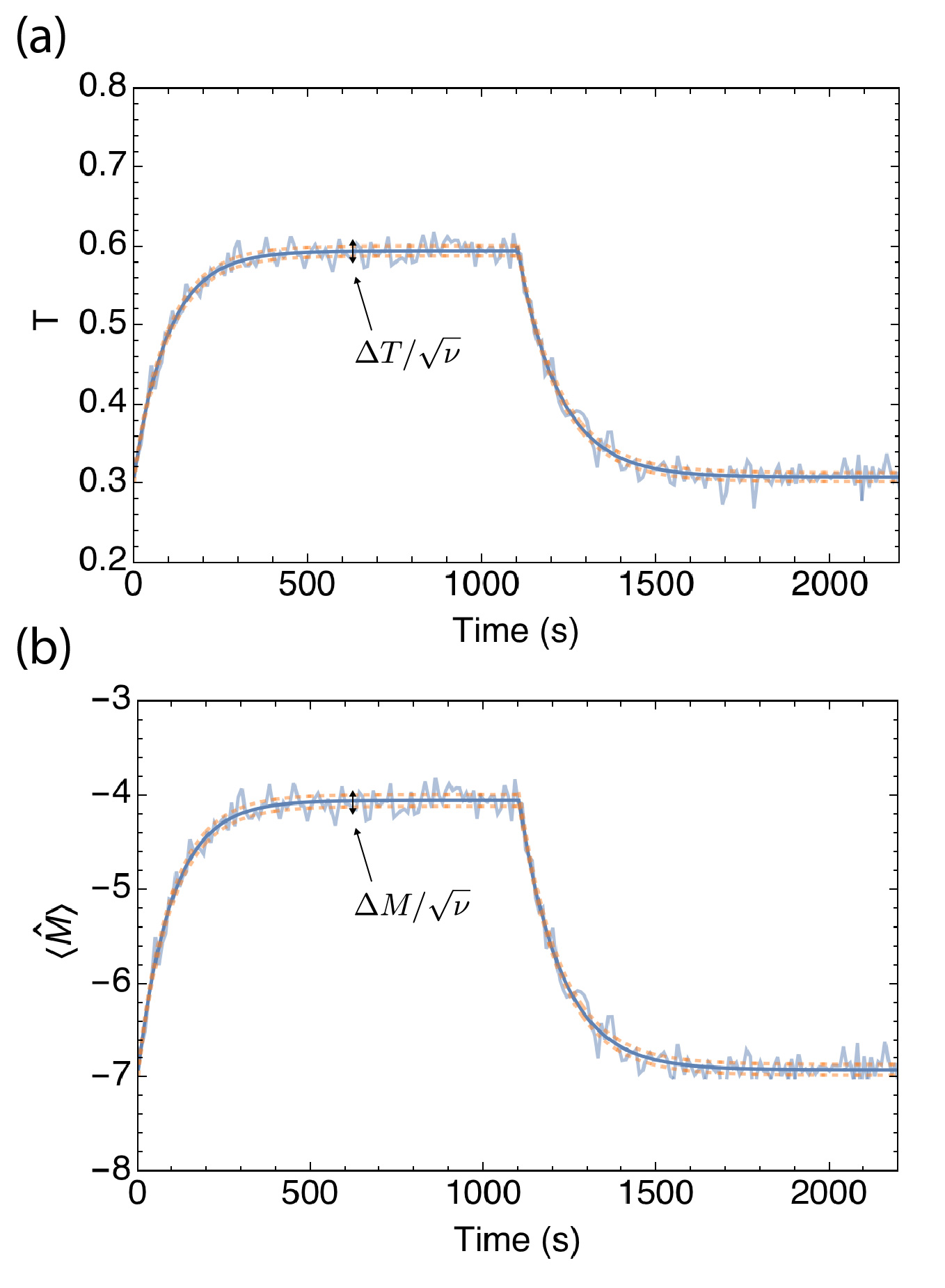}
\caption{Example sensorgrams with noise (shot noise) using the classical TMC state. (a) The mean transmittance $T$ sensorgram. (b) The measurement outcome $\langle \hat{M} \rangle$ sensorgram. In both sensorgrams the ideal mean value is shown as a solid blue line and the simulated sensorgram with noise is shown as a solid light blue line ($\bar{T}$ for transmittance and $\bar{M}$ for the intensity-difference). The upper and lower dashed orange lines give the total noise added to the ideal sensorgram, which for (a) is $\Delta T/\sqrt{\nu}$ and for (b) is $\Delta M/\sqrt{\nu}$. In both plots the parameters used are $N=10$, $\eta_a=\eta_b=1$ and $\nu=1000$. The specific sensor details, such as operating wavelength and angle, as well as the kinetic parameters that cause the variation in $T$ shown are taken from the study introduced in section IV, where more details are provided.}
\label{fig2} 
\end{figure}

The type of measurement we consider in this study is an intensity-difference measurement between modes $a$ (signal) and $b$ (reference), defined mathematically as $\langle {\hat{M}} \rangle = \langle \hat{{a}}^{\dagger} \hat{{a}} \rangle -\langle \hat{{b}}^{\dagger}\hat{{b}}  \rangle$ from which the sensitivity can be obtained. To calculate $\Delta T$ we require $\Delta M$, which can be found using the explicit formula below
\bqa
\Delta M&=&[\Delta n_a^2+\Delta n_b^2 - 2 (\langle \hat{n}_a \hat{n}_b\rangle-\langle \hat{n}_a \rangle \langle \hat{n}_b \rangle)]^{1/2}.
\label{eqn:7}
\eqa
For both $\langle {\hat{M}} \rangle$ and $\Delta M$ the expectation value is taken with respect to the final state on which the measurement is performed. We then have that the TMC, TMF and TMSV states all give $\langle {\hat{M}} \rangle=(\eta_aT-\eta_b)N$, where $N$ is the mean photon number in each of the input modes $a$ and $b$~\cite{Lee2017,Tame19}. For the TMSD state we have $\langle {\hat{M}} \rangle=\eta_aT N-\eta_b(N-|\alpha|^2)$, which is equivalent to the expectation value of the other states in the limit $\cosh^2 r\gg1$, {\it i.e.}, large squeezing. The expressions for $\Delta M$ for all the states used in this study are given in Appendix~\ref{sec:noise}.

In order to extract out the kinetic parameters of a given receptor-ligand interaction, the transmittance $T(t)$ must be measured over time. This can be obtained from the time dependence of the intensity-difference measurement, $\langle \hat{M} \rangle$, as it is linearly related to $T$ for a set of system parameters $N$, $\eta_a$ and $\eta_b$ (and $|\alpha|^2$ for the TMSD state). From the sensorgram signal $T(t)$ a nonlinear fit is then required (the details of which are given in Section III). Due to the non-zero estimation precision $\Delta T$ at a given instance of time when a measurement is performed, as a result of the uncertainty $\Delta M$, the measured sensorgram will have some associated noise and a nonlinear fit must be done that takes into account this noise. As it is not straightforward to model how the noise in the sensorgram at each instance of time/measurement translates into noise in the kinetic parameters in an analytical way, in this work we estimate the kinetic parameters and find their estimation precision by performing a Monte Carlo numerical simulation. In this simulation of a potential experiment, we start with an ideal sensorgram with a fixed temporal profile $T(t)$ and at each instance of time we consider fluctuations in $T$ according to $\Delta T$ for a given quantum state. 

To make the scenario more relevant to an experiment, we consider that at each instance of time ({\it e.g.}~every second) $\nu$ measurements are made. The assumption here is that the measurements would be performed on a time-scale much faster than the change in $T$. Thus, the fluctuations in $T$ that would be measured, {\it i.e.}, the means $\bar{T}$, are according to $\Delta T/\sqrt{\nu}$. For a given state of light, a simulated sensorgram then follows the ideal $T(t)$ but with noise applied at each instance of time according to the state, as shown in Fig.~\ref{fig2}(a) for the classical TMC state.

It is important to point out that the noise $\Delta T$ of a state like the TMF state is not Gaussian (it is binomial), however, for each instance of time we consider Gaussian noise for the mean of a set of $\nu$ measurements with standard deviation given by $\Delta T/\sqrt{\nu}$, where $\Delta T$ is the noise for the particular state. This approach is justified as we are using the sample mean as an estimator. Each instance of time in the sensorgram corresponds to fluctuations of the sample mean (where a sample is made up of $\nu$ measured values in a set) which follows a Gaussian distribution regardless of the underlying probability distribution for a state according to the central limit theorem~\cite{Hogg2013}.

A single simulated noisy sensorgram, such as that shown in Fig.~\ref{fig2}(a), will however only give a single value for a kinetic parameter $k$ from a nonlinear fit. In order to find an estimation of the mean of a parameter and its precision we require more than one sensorgram. We therefore simulate $m$ sensorgrams and from these we obtain a sample mean of $\bar{k}$ for a given kinetic parameter. We then repeat this sampling $p$ times in order to build up a distribution of $\bar{k}$'s, which has a mean and standard deviation that are stable as $p$ increases. The mean of the $\bar{k}$'s of this distribution is then the estimation and the standard deviation $\Delta \bar{k}$ is the estimation precision. Practically what this means is that we have the estimation precision of the kinetic parameter $k$ for a single set of $m$ sensorgrams, each of which consists of $\nu$ measurements at each instance of time. In this sense, the parameter $m$ plays the role of a set in the same way $\nu$ does for a set of measurements at an instance of time in the sensorgram and the estimation precision $\Delta \bar{k}$ is expected to scale as $1/\sqrt{\nu m}$ for arbitrary $\nu$ and $m$. We study if this is indeed the case. 

The parameters $\nu$ and $m$ are separated explicitly in this work as this gives the ability to improve the estimation precision for a fixed $\nu$ by increasing the number of sensorgrams in a set, {\it i.e.}, $m$. An upper bound on the value of $\nu$ is set by the temporal profile of the sensorgram, where it is assumed that one can measure a set of $\nu$ probe states over a small finite duration around a given instance in time, for which $T$ remains roughly constant due to the slowly varying temporal profile of the sensorgram. The assumption applies to most cases where the change of $T$ is slow compared to the time window of a given detector. When the assumption is not valid, one could also think about doing a smaller number of measurements per instance of time by reducing $\nu$. In this case, the signal-to-noise ratio per instance of time is decreased due to the reduced sample size $\nu$, and so $m$ can be used to improve the resulting estimation precision by including more sensorgrams in a set. In the case of a fixed $m$ value, {\it e.g.}, $m=1$, the estimation precision of the kinetic parameters has only a $\nu$ dependence.

It is also important to note that it may be desirable from an experimental point-of-view for the $m$ independent sensorgrams to all be taken from a single sensorgram `run'. This is because performing $m$ identical experiments can be challenging due to the variation in the preparation of the analyte, resetting the sensor and temperature fluctuations amongst other factors. While these things would add additional noise to the precision and we do not consider them here, we simply point out that the $m$ sensorgrams could be taken from a single sensorgram by sampling $m$ sets of $\nu$ measurements at each instance of time if the measurements are performed fast enough. The equivalence between $m$ independent sensorgrams and $m$ samplings at a given instance of time for a single sensorgram is valid as the sensorgram can be considered as an ergodic process~\cite{Peebles2001,Porat1994}. In this case, at a fixed point in time, $t_0$, we have that $T(t_0+\Delta t)$ is a wide-sense stationary process over the interval $\Delta t$ as $\Delta t \to 0$. This is because the noise in the mean transmittance obtained from a set of $\nu$ measurements at any point in time from $t_0$ to $t_0+\Delta t$ is Gaussian regardless of the underlying noise model for each of the $\nu$ measurements. Such behavior could be checked in an experiment using the augmented Dickey-Fuller test statistic~\cite{Fuller1976}.

Finally, instead of extracting the kinetic parameters from the sensorgram for $T$ with noise $\Delta T$ we use the sensorgram obtained directly from the measurement $\langle \hat{M}\rangle$ and associated noise $\Delta M$ because, as mentioned already, both are linearly related to their $T$ counterparts, {\it i.e.}, $\langle {\hat{M}} \rangle=\eta_aNT-\eta_bN$ and $\Delta M=\eta_a N \Delta T$. In other words, one can extract the kinetic parameters either from the pair $(T,\Delta T)$ or from the pair $(\langle {\hat{M}} \rangle,\Delta M)$. We have chosen the latter for convenience and in Fig.~\ref{fig2}(b) we show the corresponding sensorgram for this pair, which can be compared with the sensorgram for the pair $(T,\Delta T)$ shown in Fig.~\ref{fig2}(a).

\subsection{The three sensing scenarios}

The general sensing model considered so far and shown in Fig.~\ref{fig1}(c) is a `standard two-mode sensing' scenario where we set the loss in either mode to be the same, {\it i.e.}, $\eta_a = \eta_b = 1$ in the ideal case when we disregard loss in the system, or in a realistic case when including some loss in both modes, {\it e.g.}, $\eta_a = \eta_b =0.8$. Reaching this value of loss is certainly challenging in an experiment, but not out of reach -- the loss in an experiment would be mainly due to the intrinsic detector efficiency of commonly used detectors in quantum experiments ($\sim$0.6)~\cite{Lee2018,Peng2020,Zhao2020}, but it is also a result of the overall transmission drop for light passing through the prism due to the optical density of the prism’s material. If a small enough prism is used (or a plasmonic waveguide is substituted instead -- see Refs.~\cite{Peng2020,Zhao2020}) and state-of-the-art detectors employed~\cite{Dauler2014}, then $\eta=0.8$ is a realistic loss value that can be attained in experiments. 

From Refs.~\cite{Tame19,Lee2021} we also know that by varying the values of the losses in the different modes, namely $\eta_b = \eta_aT$, we can gain a further reduction in the photon-number differential noise in the measurement for some of the states. This can be considered to be a form of optimization of our sensing model. This second scenario, which we call `optimized two-mode sensing' is also included in our study. 

Finally, we go one more step further and reduce the two-mode sensing model to a single-mode model by removing the reference mode $b$ and return to the conventional model for plasmonic sensing. We can effectively achieve this by maximizing the loss in mode $b$, {\it i.e.}, setting $\eta_b = 0$ such that there will be no transmittance in that mode and the intensity-difference measurement is simply an intensity measurement. We call this third scenario `single-mode sensing'.

\section{Interaction kinetics}

Interaction kinetics refers to the dynamic binding and unbinding processes of ligands to receptors~\cite{Xiao19}, which are divided into 3 main phases: association, steady state and dissociation, as shown in Fig.~\ref{fig1}(b). The association phase refers to the binding of ligands to receptors to form receptor-ligand complexes, and the steady state phase is where equilibrium is reached as the number of ligands binding equals the number which are unbinding. Finally, the dissociation phase is the breaking of bonds between the ligands and receptors. In Appendix~\ref{sec:intkin} we provide details of the model we use for the interaction kinetics~\cite{Xiao19} and how it is linked to the transmittance, $T$, of the sensor. Below is a brief summary of the model.

The link between the concentration of the receptor-ligand complex $[C]$ and the transmittance of the sensor $T$ is given by the refractive index, $\epsilon_a$, of the region above the gold surface, whose change is induced by the sequence of analytes being passed over the flow cell: (i) buffer and ligands (association and steady state) for $0 \leq t < \tau$ and (ii) buffer only (dissociation) for $t\geq \tau$. The refractive index change can be understood as a change in the dipole moments of the immobilized receptors as they are converted into complexes and then unconverted~\cite{Xiao19}. For a fixed incidence angle of light, an increase in the complex concentration $[C]$ therefore increases the value of the refractive index, $\epsilon_a$, and thus $T$, as shown in the inset of Fig.~\ref{fig1}(a). In the ideal case, when there is a linear relation between $[C]$ and $T$ we can write~\cite{Xiao19} 
\begin{equation}
T(t) =
\begin{cases}
                                  {T}_{\infty}(1-{e}^{-{k_s t}})& \text{$0 \leq t < \tau $} \\
                                   {T}_{\tau} {e}^{-{k_d (t-\tau)}} & \text{$t\geq \tau$}, \\
\end{cases}
\label{sensorgrameqm}
\end{equation}
where $T_\infty$ is a constant determined by the initial concentration of the ligands and receptors, the thickness of the receptor and ligand layers above the gold surface, and the affinity $k_A=\frac{{k_a}}{{k_d}}$. We then have the constant $T_\tau=T_\infty(1-e^{-k_s \tau})$. In the above, the kinetic parameter ${k_s}={k_a}[{L_0}]+ {k_d}$ represents the observable rate for the association phase, ${k_a}$ is the association constant measured in ${\rm M}^{-1} {\rm s^{-1}}$ (per molarity per second), ${k_d}$ is the dissociation constant measured in ${\rm s^{-1}}$ and $[{L_0}]$ is the initial ligand concentration. Equation~\eqref{sensorgrameqm} is the theoretical model for the sensor's response, which is the sensorgram that would be measured in an ideal experiment (no noise). From the measured sensorgram a nonlinear fit is then performed, {\it e.g.}, using the Gauss-Newton method, with respect to the theoretical model in order to extract out the association and dissociation kinetic parameters. From the fit, $k_d$ and $k_s$ are obtained and with a knowledge of the initial ligand concentration $[L_0]$, the parameter $k_a$ can be found from the relation ${k_a}=({k_s}-k_d)/ [{L_0}]$. 

While other methods for extracting the kinetic parameters are possible, we have chosen this method as it is one of the most direct~\cite{Xiao19}. When considering a realistic sensorgram that is measured in an experiment with noise, the extracted kinetic parameters from the fit will have the noise imparted to them and the estimate obtained will have an estimation error (estimation precision). This is the central question we seek to answer in this work -- How does the noise from different quantum states affect the noise of the extracted kinetic parameters?


\section{Large sensorgram deviation}

We start by studying the interaction of the antigen BSA interacting with the antibody rabbit anti-cow albumin IgG1 (anti-BSA), which is well documented in the work of Kausaite {\it et al.}~\cite{Kausaite07}. It has been chosen as it provides a large sensor response, or deviation, due to the large change in the refractive index during the interaction dynamics. It acts as a basic starting point for our study of quantum states being used to improve the precision in the estimation of kinetic parameters. 
\begin{figure}[t]
\centering
\includegraphics[width=8cm]{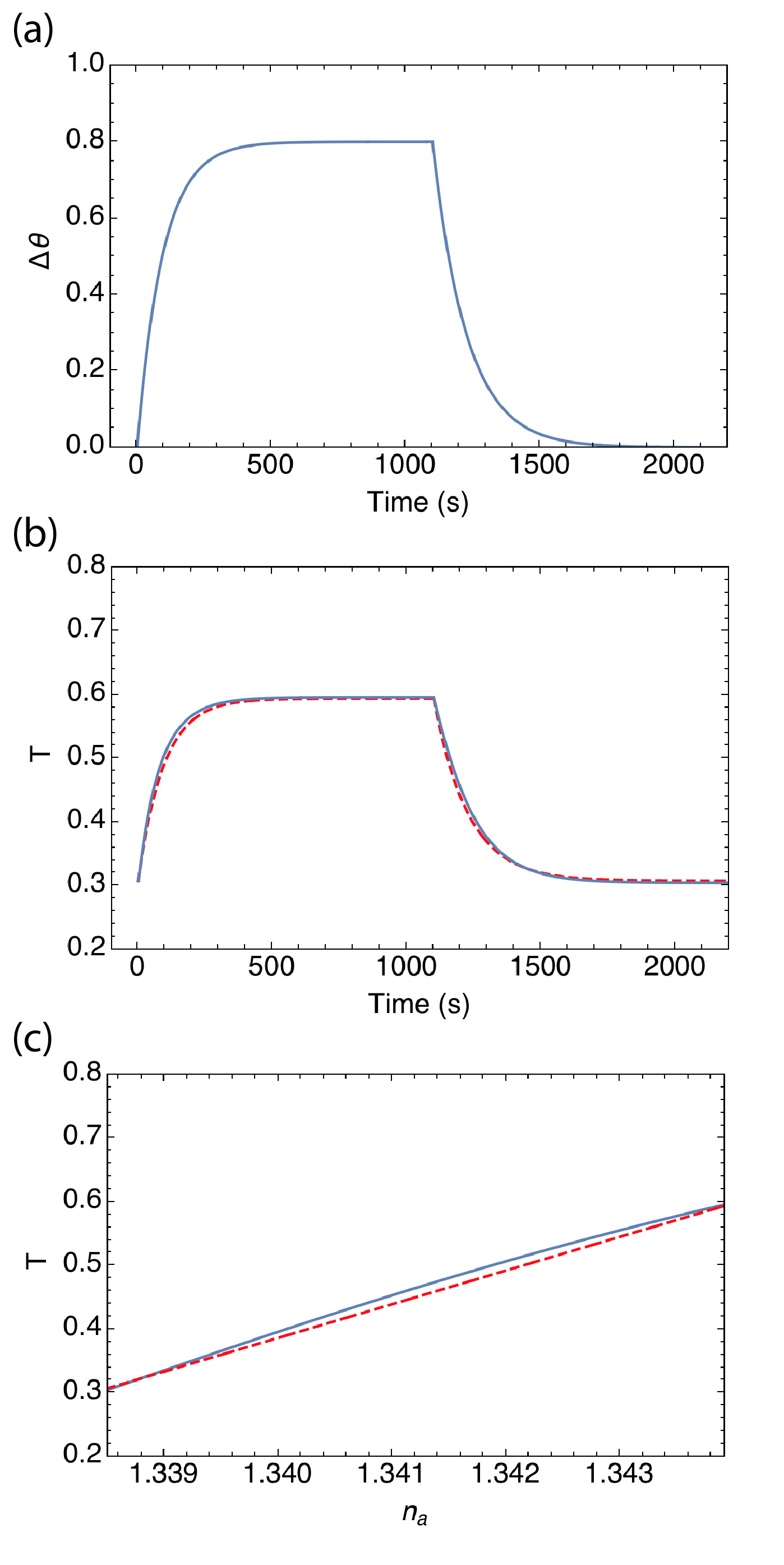}
\caption{Sensorgrams from the experiment by Kausaite {\it et al.}~\cite{Kausaite07} which investigates BSA interacting with the antibody rabbit anti-cow albumin IgG1 (anti-BSA). (a) Angular sensorgram, $\Delta\theta(t)$, where the full sensorgram is $\theta(t)=\theta(0)+\Delta\theta(t)$ and $\theta(0)=71.0966$ degrees. (b) Reconstructed transmittance sensorgram, $T(t)$ (solid line) and linearized reconstructed transmittance sensorgram, $T_L(t)$ (dashed line). For both transmittance sensorgrams, $\theta_{in}=70.1200$ degrees has been set. (c) Comparison of the nonlinear response of $T(t)$ (solid line) and linear response $T_L(t)$ (dashed line).}
\label{angtranssensorgram} 
\end{figure}

\begin{figure*}[t]
\centering
\includegraphics[width=16cm]{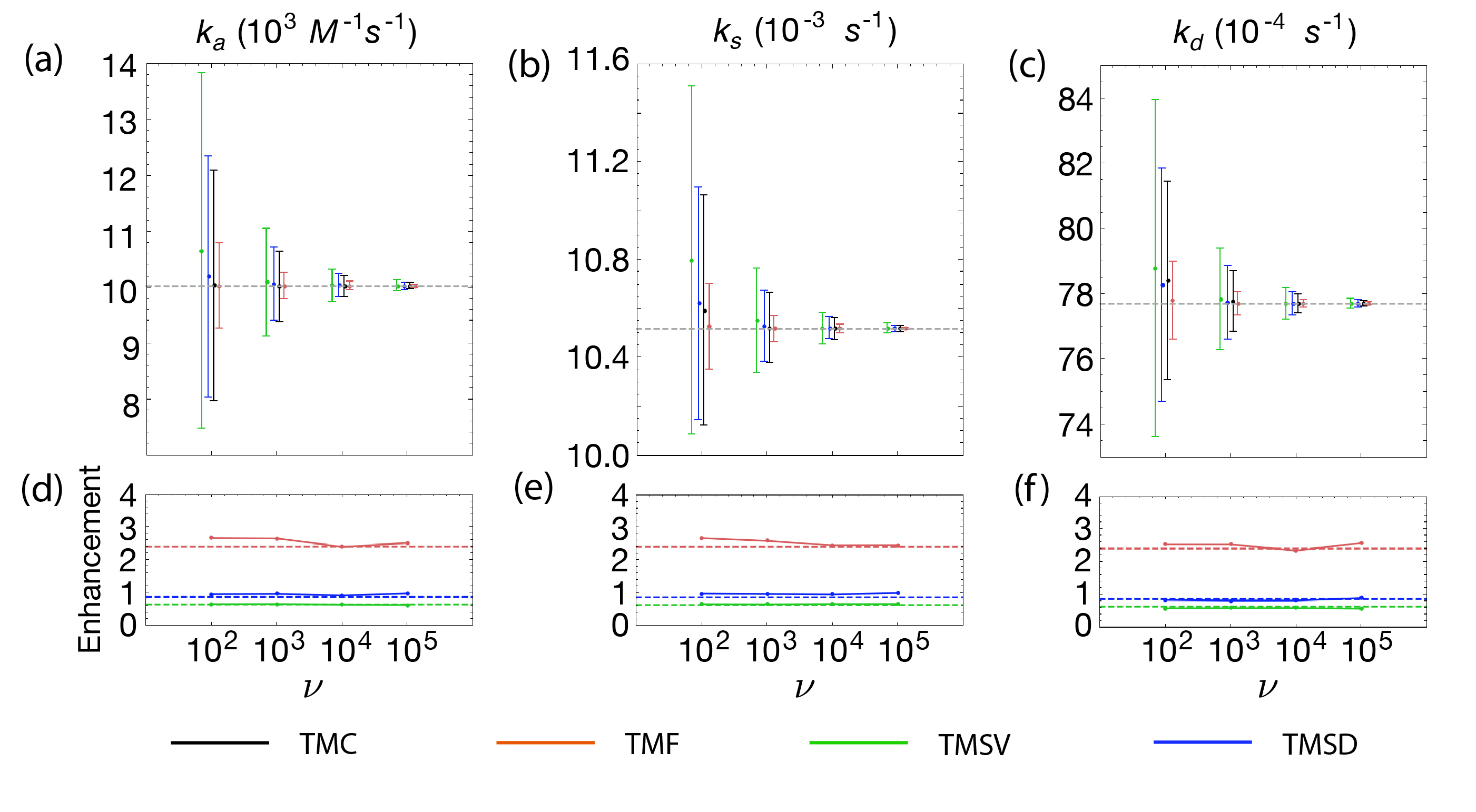}
\caption{Standard two-mode sensing using different quantum states (no loss, $\eta_a=\eta_b=1$). Panels (a), (b) and (c) show the estimation values and precisions for the kinetic parameters $k_a$, $k_s$ and $k_d$ as $\nu$ increases for $m=10$. For each value of $\nu$ the error bars represent the TMSV, TMSD, TMC and TMF states, going from left to right. Panels (d), (e) and (f) show the corresponding enhancement ratio for the different quantum states for $m=10$. From top to bottom the lines correspond to TMF, TMSD and TMSV, respectively. The dotted lines are a guide representing the enhancement expected from the ratio $R_M$ at the mid-point of the sensorgram for the respective state.}
\label{standardtwomodeKausaite} 
\end{figure*}

\subsection{Transmittance sensorgram}

In the experiment reported by Kausaite {\it et al.}, BSA proteins are immobilized on a gold surface using a self-assembled monolayer and act as the receptors. The real-time SPR curve is measured for the BSA interacting with anti-BSA in an analyte using an Autolab ESPRIT commercial SPR sensor developed by ECO Chemie~\cite{ECOChemie}. From the sensorgram obtained they extracted out the following kinetic parameters, ${k_a = 9.36 \times 10^3~{\rm M}^{-1} {\rm s}^{-1}}$, ${k_d = 7.85 \times 10^{-3}~{\rm s}^{-1}}$ and ${L_0 = 274 \times 10^{-9}~{\rm M}}$. However, in the experiment angular interrogation was used and therefore the sensorgram obtained was angle dependent. Our model for comparing quantum states is based on intensity interrogation, which is equivalent to angular interrogation in terms of sensitivity performance~\cite{Piliarik2009}, but a transformation is required to go from the angular sensorgram measured in the experiment to the corresponding intensity (transmittance) sensorgram that would be measured and we can use to compare the different quantum states. 

In order to perform the transformation we need to know what the value of $T_\infty$ (and therefore $T_\tau$) in Eq.~\eqref{sensorgrameqm} would be in the experiment. With a knowledge of this parameter, as well as the extracted kinetic parameters stated above we then have the equivalent transmittance sensorgram that we can use in our simulations. To obtain the value of $T_\infty$, we use the angular sensorgram from the experiment, which is shown in Fig.~\ref{angtranssensorgram}(a), to find the time dependence of the refractive index, $n_a(t)$, above the gold surface. Using this and the other physical parameters from the experiment we reconstruct a time dependent model $T(t)$ for the transmittance with the correct $T_\infty$ value. The details of the physical parameters and the procedure are given in Appendix~\ref{sec:Tsensor}.

With the full time dependence of $n_a(t)$ known from the angular sensorgram, we use it in Eq.~\eqref{eqn:r} to obtain $T(t)=|r_{spp}(t)|^2$, which is shown in Fig.~\ref{angtranssensorgram}(b) as a solid line. In this plot we have set $\theta_{in}=70.1200$ degrees, which corresponds to an angle close to the inflection point for the transmittance curve, as shown as a dot in the inset of Fig.~\ref{fig1}(a).

A final step in the reconstruction of the transmittance sensorgram is to check that $T(t)$ has a linear response to changes in $n_a$. The angular dependence follows closely a linear response to changes in $n_a$ and if the response of $T(t)$ is not linear, then our model for $T(t)$ will not be consistent with the angular model and it will lead to different kinetic parameters being extracted from the fits. In Fig.~\ref{angtranssensorgram}(c) we show the response of $T$ to changes in $n_a$ as the solid line and a linear model,
\be
T_L(t)=T(0)+\frac{(T(\tau)-T(0))}{(n_a^2(\tau)-n_a^2(0))}(n_a^2(t)-n_a^2(0)),
\ee
as a dashed line. One can see that the response of the transmittance $T$ is slightly nonlinear. The solution to this problem is to calibrate the sensor by sending in an analyte with a range of known refractive indicies and measuring the transmittance response. As the response of $T$ to $n_a$ is monotonic (see Fig.~\ref{angtranssensorgram}(c)), it means that for a given value of $T$ there is a corresponding value for the linear calibrated $T_L$. The correction factor is therefore transmittance dependent, $C(T)=T_L/T$, and its form is known after calibration. We then have that $T_L=C(T)T$. The linearized sensorgram is shown in Fig.~\ref{angtranssensorgram}(b) as a dashed line and this is the sensorgram we use for comparing the different quantum states, as it will give approximately the same kinetic parameters as the $\theta(t)$ sensorgram in the ideal case when there is no noise. The $T(t)$ and $T_L(t)$ sensorgrams are very similar to each other, which is due to the small non-linear response that nonetheless needs to be accounted for.

The kinetic parameters obtained from the ideal $T_L(t)$ sensorgram shown as the dashed line in Fig.~\ref{angtranssensorgram}(b) are $k_s=0.0105 ~{\rm s}^{-1}$, ${k_d = 7.771 \times 10^{-3}~{\rm s}^{-1}}$ and ${k_a = 10.029 \times 10^3~{\rm M}^{-1} \rm{s}^{-1}}$, where we have used ${L_0 = 274 \times 10^{-9}~{\rm M}}$ to extract $k_a$ from the parameters $k_s$ and $k_d$ using the formula $k_s=k_a L_0+k_d$. These parameters are slightly different in value to those extracted from the angular sensorgram in the experiment due to a small residual nonlinear response of the corresponding transmittance sensorgram after the linearization. However, we use these values as the ideal values in our study as they correspond to the ideal transmittance sensorgram that we have obtained as our model. We now consider using the classical TMC state, and compare the estimate and precision of the kinetic parameters obtained with it to those obtained using the different quantum states. To do this we simulate the measurement process and noise according to the Monte Carlo simulation method described in Section II C.

\subsection{Standard two-mode sensing}

In this first scenario, we consider the general sensing model shown in Fig.~\ref{fig1}(c), where we set the loss in either mode to be the same. To start with, we take the ideal case of no loss, {\it i.e.}, $\eta_a = \eta_b = 1$. In Fig.~\ref{standardtwomodeKausaite}(a)-(c) we show the estimation value (mean as a point) and estimation precision (standard deviation as an error bar) for the kinetic parameters $k_a$, $k_s$ and $k_d$ for increasing sample size $\nu$. For this example, we have used $N=10$ for the photon number and $m=10$ for the number of sensorgrams in a set. The number of sets of sensorgrams simulated is $p=1500$, which is chosen as it provides a stable distribution of the extracted kinetic parameters from the fits. We use this value of $p$ for all the simulations in this work. 

One can see in Fig.~\ref{standardtwomodeKausaite}(a)-(c) that the TMF state provides the best estimation of the kinetic parameters for any $\nu$, followed by the TMC state, then the TMSD state and finally the TMSV state. The estimation precision shown for each state physically corresponds to that of a fixed set of $m=10$ sensorgrams, each of which has $\nu$ states probed at a given instance of time, with a step-size between instances of time of 10s, as described in more detail in Section II C. In the present scenario, the sensorgram is 2200 seconds in duration and so there are 220 points in total, each point having $\nu$ probe states measured. The step-size for the points was chosen so that there was a fine enough mesh for the fit to return the exact values in the ideal case when there is no noise.

We quantify the improvement in the estimation precision by considering the ratio, $R_k$, of a quantum state's measurement precision for parameter $k$, given by $\Delta k_Q$, to that of the TMC state with matching mean photon number in each mode, $\Delta k_C$, {\it i.e.}, $R_k=\Delta k_C/\Delta k_Q$. We call this the enhancement ratio. The enhancement is shown in Fig.~\ref{standardtwomodeKausaite}(d)-(f) for $k_a$, $k_s$ and $k_d$ as $\nu$ increases. The TMC state has no enhancement and the ratio value is 1 naturally. 

In Fig.~\ref{standardtwomodeKausaite}(d)-(f), the dotted lines for each state are a guide that represents the enhancement expected from the ratio of $R_M=\Delta M_C/\Delta M_Q$ at the mid-point value of the sensorgram (at $T=0.451$), as shown in Fig.~\ref{angtranssensorgram}(b). The ratio $R_M$ is related to the well known noise reduction factor (NRF) used to quantify how well quantum states reduce measurement noise by the relation $R_M=1/\sqrt{NRF}$~\cite{Jedrkiewicz2004,Bondani2007,Blanchet2008,Perina2012,Lee2021}. The enhancements $R_k$ are clearly roughly in line with that expected from $R_M$ at the mid-point value of $T$. It is interesting that the enhancement of the estimation precision for the kinetic parameters can be found from assessing only $R_M$ (or the NRF) for the mid-point value of the sensorgram. This clearly shows that the enhancement in the estimation precision of a parameter extracted from a static transmittance, as studied in Ref.~\cite{Lee2017}, carries over to parameters extracted from a dynamic transmittance, with the enhancement approximated well by the mid-point enhancement of the dynamic transmittance. In Appendix~\ref{sec:midpointenhance} we show how the enhancement $R_M$ changes about the mid-point of the sensorgram for each of the states. The general trend is that $T$ values below (above) the mid-point give a lower (higher) enhancement. The overall effect over the range of $T$ in a sensorgram appears to be an averaging of the enhancement about the mid-point, which gives approximately the enhancement $R_k$ for the kinetic parameters.
\begin{figure*}[t]
\centering
\includegraphics[width=16cm]{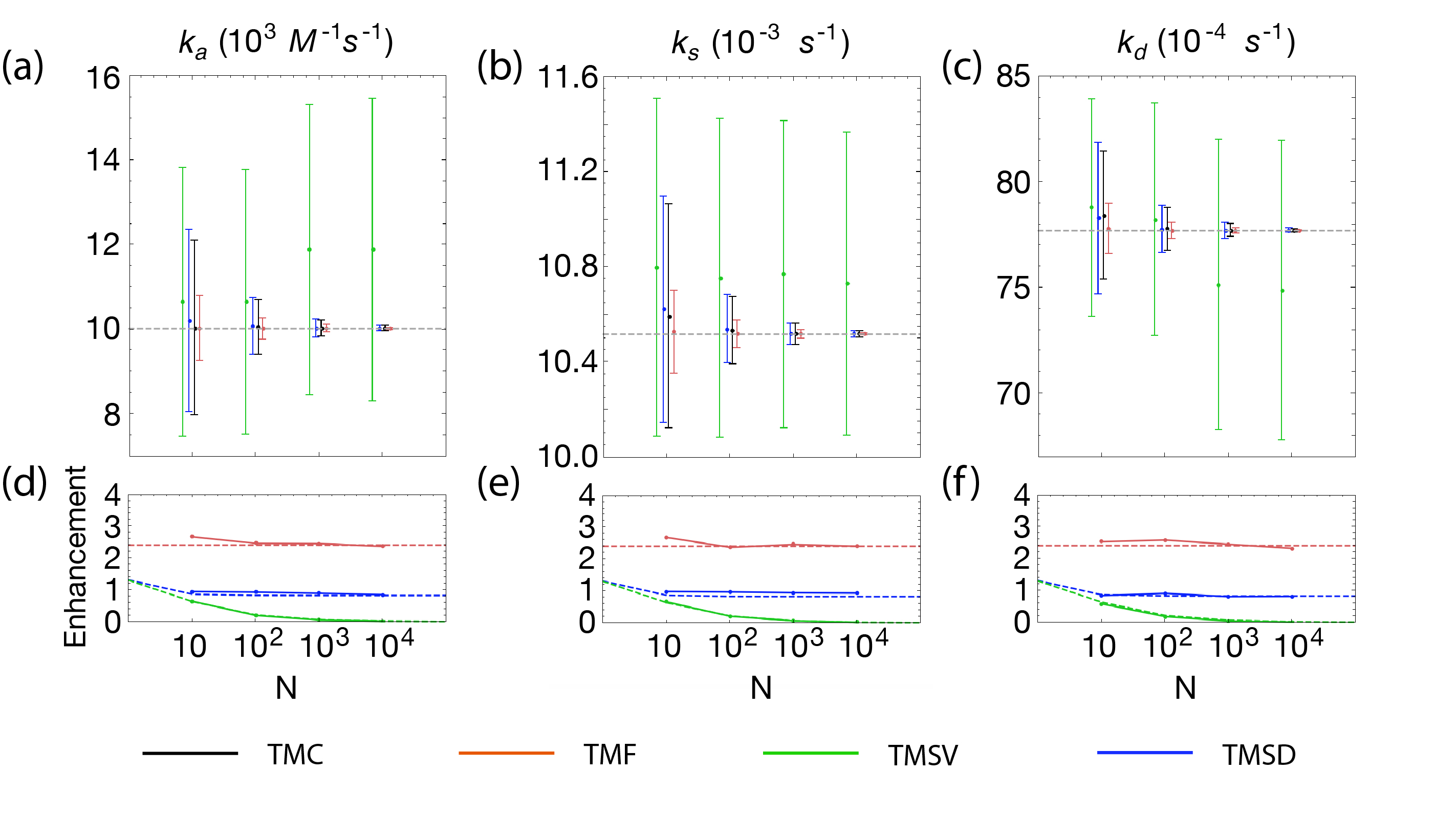}
\caption{Photon number dependence in standard two-mode sensing using different quantum states (no loss, $\eta_a=\eta_b=1$). Panels (a), (b) and (c) show the estimation values and precisions for the kinetic parameters $k_a$, $k_s$ and $k_d$ as $N$ increases for $m=10$ and $\nu=100$. For each value of $N$ the error bars represent the TMSV, TMSD, TMC and TMF states, going from left to right. Panels (d), (e) and (f) show the corresponding enhancement ratio for the different quantum states. From top to bottom the lines correspond to TMF, TMSD and TMSV, respectively. The dotted lines for each state are a guide representing the enhancement expected from the ratio $R_M$ at the mid-point of the sensorgram for the respective state.}
\label{standardtwomodeKausaitephoton} 
\end{figure*}

In Appendix~\ref{sec:enhancem} we consider the effect that changing $m$ (the number of sensorgrams in a set) has on the enhancement. By setting $m=50$, a similar behavior to that shown in Fig.~\ref{standardtwomodeKausaite} for $m=10$ can be seen for the estimation precision of all the states, with the TMF state providing the best estimation in the kinetic parameters for any $\nu$, followed by the TMC state, then the TMSD state and finally the TMSV state. We call the ratio of the enhancement ratio for $m=50$ and $m'=10$, {\it i.e.}, $R_{k,50}/R_{k,10}$, for $k_a$, $k_s$ and $k_d$, the `$m$-enhancement' ratio. It is expected to be $\sqrt{m/m'}=2.236$ due to the $1/\sqrt{m}$ dependence of the estimation precision, $\Delta k$, for a fixed $\nu$. As can be seen in Appendix~\ref{sec:enhancem}, the $m$-enhancement ratios for the different states are roughly in line with the expected value 2.236 when going from $m=10$ to $m=50$ sensorgrams in a set. 

\begin{figure*}[t]
\centering
\includegraphics[width=18cm]{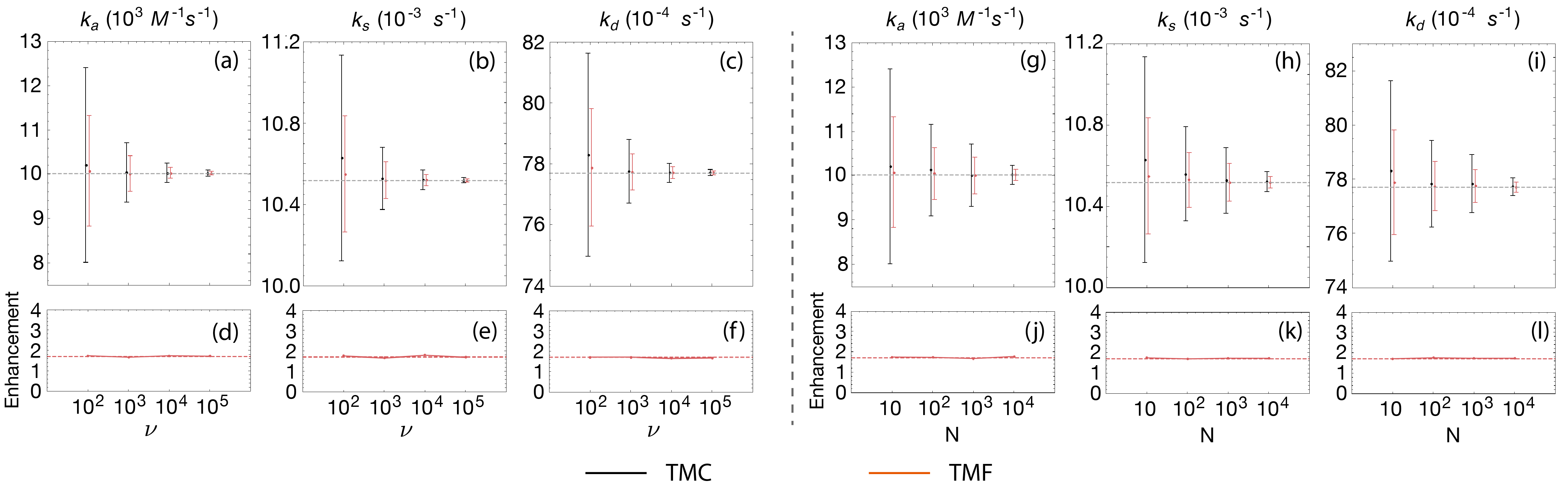}
\caption{Standard two-mode sensing using the TMF and TMC quantum states (loss, $\eta_a=\eta_b=0.8$). Panels (a), (b) and (c) show the estimation values and precisions for the kinetic parameters $k_a$, $k_s$ and $k_d$ as $\nu$ increases for $m=10$. For each value of $\nu$ the error bars represent the TMC and TMF states, going from left to right. Panels (d), (e) and (f) show the corresponding enhancement ratio for the TMF state for $m=10$. Panels (g), (h) and (i) show the estimation values and precisions for the kinetic parameters $k_a$, $k_s$ and $k_d$ as $N$ increases for $m=10$ and $\nu=100$. For each value of $N$ the error bars represent the TMC and TMF states, going from left to right. Panels (j), (k) and (l) show the corresponding enhancement ratio for the TMF state. In panels (d), (e), (f), (j), (k) and (l) the dotted lines are a guide representing the enhancement expected from the ratio $R_M$ at the mid-point of the sensorgram for the TMF state.}
\label{standardtwomodeKausaiteloss} 
\end{figure*}

In addition to checking how well quantum states of a fixed photon number ($N=10$) enhance the estimation precision as $\nu$ changes, we also study the dependence of the photon number for a fixed $\nu$. In this case, we set $\nu=100$ and vary $N$ from 10 to 10,000, as shown in Fig.~\ref{standardtwomodeKausaitephoton}(a), (b) and (c) for $k_a$, $k_s$ and $k_d$, respectively. One can see that as $N$ increases the TMF state provides the best estimation in the kinetic parameters for any $N$, followed by the TMC state, then the TMSD state and finally the TMSV state. Interestingly, the TMSV state has a photon number dependence, which is a known behavior in the static case~\cite{Lee2017,Lee2021}. The estimation precision becomes worse as $N$ increases and the accuracy (deviation of the estimation value from the ideal value) of the estimation also does not improve. This can be seen more clearly in the enhancement plots in Fig.~\ref{standardtwomodeKausaitephoton}(d), (e) and (f) and anticipated from the form of $R_M=\Delta M_{TMC}/\Delta M_{TMSV}$ using the relations for the $\Delta M$'s given in Appendix~\ref{sec:noise}. In this standard two-mode sensing scenario, the TMF state is clearly the state that offers the best estimation precision and unbiased estimation value, providing an enhancement over the classical TMC state. In Appendix~\ref{sec:midpointenhance} we show how the enhancement $R_M$ changes about the mid-point of the sensorgram for each of the states as $N$ increases. As before, the general trend is that $T$ values below (above) the mid-point give a lower (higher) enhancement. The overall effect on the kinetic enhancement $R_k$ is an averaging of the $R_M$ enhancement about the mid-point.
\begin{figure*}[t]
\centering
\includegraphics[width=18cm]{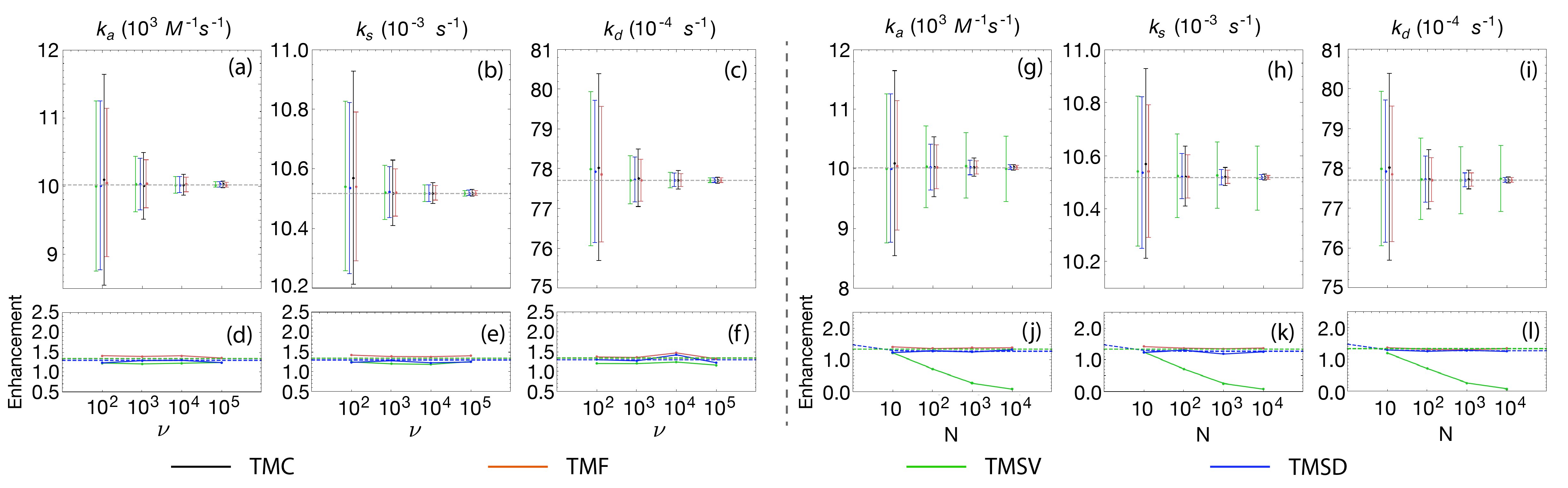}
\caption{Optimized two-mode sensing using different quantum states (no loss, $\eta_a=1$ and $\eta_b=\eta_aT_{mid}$). Panels (a), (b) and (c) show the estimation values and precisions for the kinetic parameters $k_a$, $k_s$ and $k_d$ as $\nu$ increases for $m=10$. For each value of $\nu$ the error bars represent the TMSV, TMSD, TMC and TMF states, going from left to right. Panels (d), (e) and (f) show the corresponding enhancement ratio for the different quantum states for $m=10$. From top to bottom the lines correspond to TMF, TMSD and TMSV, respectively. Panels (g), (h) and (i) show the estimation values and precisions for the kinetic parameters $k_a$, $k_s$ and $k_d$ as $N$ increases for $m=10$ and $\nu=100$. For each value of $N$ the error bars represent the TMSV, TMSD, TMC and TMF states, going from left to right. Panels (j), (k) and (l) show the corresponding enhancement ratio for the different quantum states. From top to bottom the lines correspond to TMF, TMSD and TMSV, respectively. In panels (d), (e), (f), (j), (k) and (l) the dotted lines are a guide representing the enhancement expected from the ratio $R_M$ at the mid-point of the sensorgram for the respective state.}
\label{optimizedtwomodeKausaite} 
\end{figure*}

Having confirmed the performance of the different quantum states as $\nu$, $m$ and $N$ vary, we now turn our attention to the impact of loss on the signal and reference modes, representing a more practical assessment in a potential experimental setting. As the TMF state outperforms the other states in the case of no loss, we focus on it and compare it with the classical TMC state. In Fig.~\ref{standardtwomodeKausaiteloss}(a)-(c) we show the estimation value and estimation precision for the kinetic parameters $k_a$, $k_s$ and $k_d$ for increasing set size $\nu$, with $\eta_a=\eta_b=0.8$. For this example, we have used $N=10$ for the photon number and $m=10$. One can see that even in the presence of moderate loss the TMF state clearly provides the best estimation in the kinetic parameters for any $\nu$.

The enhancement is shown in Fig.~\ref{standardtwomodeKausaiteloss}(d)-(f) for $k_a$, $k_s$ and $k_d$ as $\nu$ increases. As before, the dotted lines are a guide that represent the enhancement expected from the ratio $R_M$ at the mid-point of the sensorgram. The enhancements are again roughly in line with that expected from the mid-point value and independent of $\nu$. In Appendix~\ref{sec:midpointenhance} we show how the enhancement changes about the mid-point of the sensorgram for each of the states when there is loss.

In Fig.~\ref{standardtwomodeKausaiteloss}(g)-(i) we show the dependence of the estimation values and precisions on the photon number for a fixed $\nu$. In this case, we set $\nu=100$ and vary $N$ from 10 to 10,000. As $N$ increases the TMF state again provides the best estimation in the kinetic parameters for any $N$. In Fig.~\ref{standardtwomodeKausaiteloss}(j)-(l) we show the corresponding enhancement behavior. 

\subsection{Optimized two-mode sensing}

In a second scenario, we follow Refs.~\cite{Tame19,Lee2021} and set $\eta_b = \eta_aT$ in order to gain a further reduction in the overall noise in the measurement for some of the states. This is a form of optimization for the sensing model and helps the TMSD and TMSV states in particular in the static case. However, in the dynamic case, in order to set $\eta_b = \eta_aT$ in the reference mode we must know $T$ at each instance of time. This is not practical from an experimental point-of-view and therefore we choose $T$ to be fixed in the reference mode at the mid-point value $T_{mid}=0.4507$, {\it i.e.}, $\eta_b=\eta_a T_{mid}$, while $T$ varies in the signal mode. This is motivated by the observation in the previous section that the overall enhancement is approximately the value found at the mid-point of the sensorgram. The question we seek to answer here is whether the enhancement in the estimation precision of a parameter extracted from a static transmittance carries over to parameters extracted from a dynamic transmittance when an optimization is performed.
\begin{figure*}[t]
\centering
\includegraphics[width=18cm]{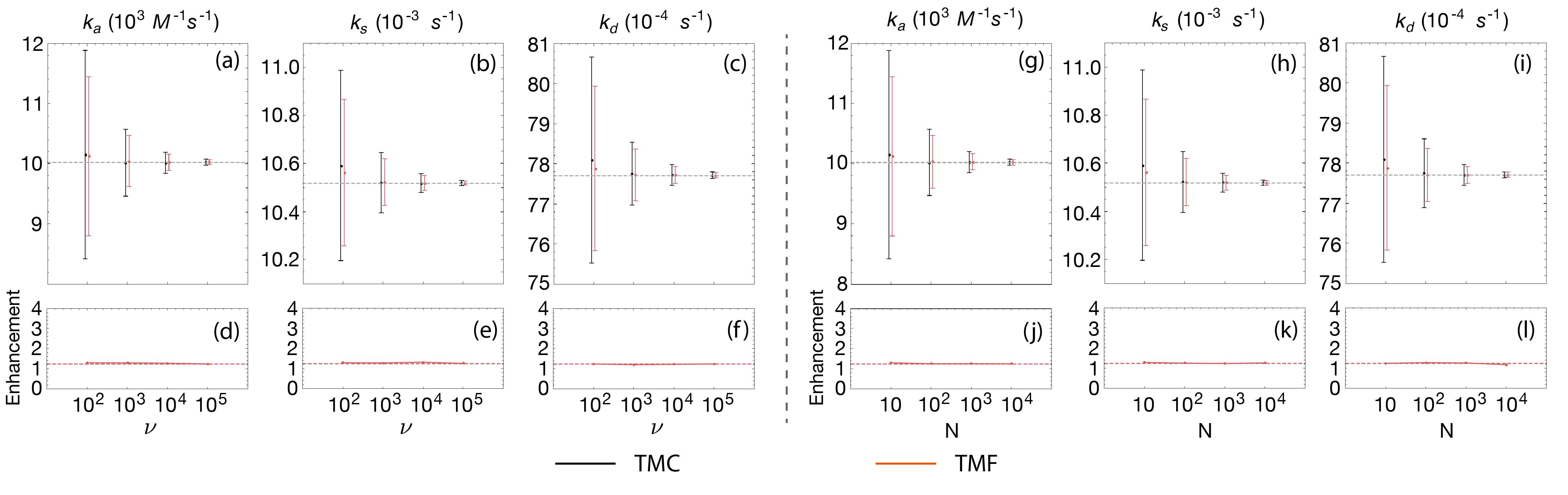}
\caption{Optimized two-mode sensing using the TMF and TMC quantum states (loss, $\eta_a=0.8$ and $\eta_b=\eta_aT_{mid}$). Panels (a), (b) and (c) show the estimation values and precisions for the kinetic parameters $k_a$, $k_s$ and $k_d$ as $\nu$ increases for $m=10$. For each value of $\nu$ the error bars represent the TMC and TMF states, going from left to right. Panels (d), (e) and (f) show the corresponding enhancement ratio for the TMF state for $m=10$. Panels (g), (h) and (i) show the estimation values and precisions for the kinetic parameters $k_a$, $k_s$ and $k_d$ as $N$ increases for $m=10$ and $\nu=100$. For each value of $N$ the error bars represent the TMC and TMF states, going from left to right. Panels (j), (k) and (l) show the corresponding enhancement ratio for the TMF state. In panels (d), (e), (f), (j), (k) and (l) the dotted lines are a guide representing the enhancement expected from the ratio $R_M$ at the mid-point of the sensorgram for the TMF state.}
\label{optimizedtwomodeKausaiteloss} 
\end{figure*}

In Fig.~\ref{optimizedtwomodeKausaite}(a)-(c) we show the estimation value and estimation precision for the kinetic parameters $k_a$, $k_s$ and $k_d$ for increasing $\nu$, with $N=10$ and $m=10$. As before, one can see that the TMF state provides the best estimation in the kinetic parameters for any $\nu$. However, different to the standard two-mode sensing scenario, the TMSV and TMSD states now both outperform the classical TMC state. This optimized scenario clearly helps the TMSV and TMSD states, where the choice of $\eta_b=\eta_aT$ reduces the respective $\Delta M$'s, as given in Appendix~\ref{sec:noise}. The corresponding enhancement is shown in Fig.~\ref{optimizedtwomodeKausaite}(d)-(f) for $k_a$, $k_s$ and $k_d$ as $\nu$ increases. The enhancements are all similar and roughly in line with that expected from the mid-point value (dotted line) and independent of $\nu$. The TMF state mid-point enhancement is the same as the TMSV state and thus its dotted line cannot be seen in the plots. In Appendix~\ref{sec:midpointenhance} we show how the enhancement changes about the mid-point of the sensorgram for each of the states. Unlike the standard two-mode scenario, the enhancement for $T$ values around the mid-point is roughly constant for the TMF and TMSD states. On the other hand, the enhancement for the TMSV state deviates slightly from the expected mid-point enhancement. This can be explained from the pronounced decrease in the enhancement on either side of the mid-point, as seen in Appendix~\ref{sec:midpointenhance}.

The dependence of the estimation value and precision on the photon number for a fixed $\nu$ is shown in Fig.~\ref{optimizedtwomodeKausaite}(g), (h) and (i) for $k_a$, $k_s$ and $k_d$, respectively. In this case, we set $\nu=100$ and vary $N$ from 10 to 10,000. One can see that similar to the standard case, as $N$ increases the TMF state provides the best estimation in the kinetic parameters for any $N$. However, different to the standard case the TMSD state now performs better than the classical TMC state for any $N$. The TMSV state also performs better than the TMC state for low photon number ($\leq 10$), although for larger $N$ the optimized scenario does not provide an advantage -- a behavior known from the static case~\cite{Tame19,Lee2021}. In Appendix~\ref{sec:midpointenhance} we show how the enhancement changes about the mid-point of the sensorgram for each of the states as $N$ increases. The enhancement for $T$ values around the mid-point is roughly constant for the TMF and TMSD states, but for the TMSV state it reduces sharply on either side. This is what causes the enhancement of the kinetic parameters to not match up with the expected mid-point value in Fig.~\ref{optimizedtwomodeKausaite}(j), (k) and (l). The TMF state mid-point enhancement is the same as the TMSV state and thus its dotted line cannot be seen in the plots.
\begin{figure*}[t]
\centering
\includegraphics[width=18cm]{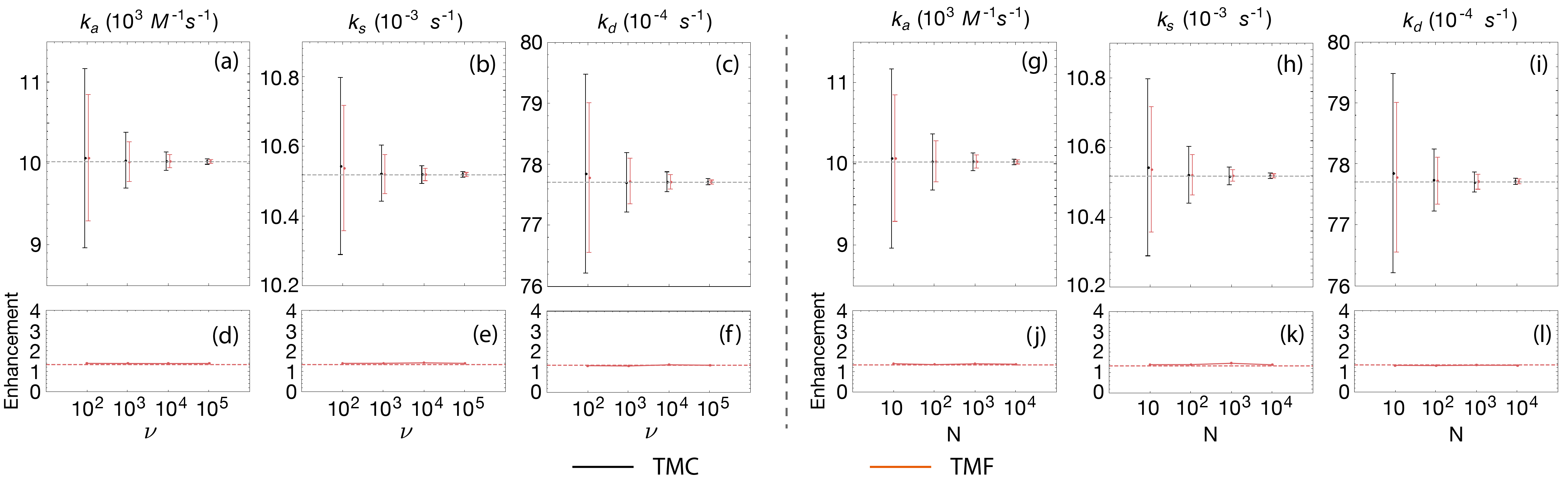}
\caption{Single-mode sensing using Fock (TMF) and coherent (TMC) states (no loss, $\eta_a=1$ and $\eta_b=0$). Panels (a), (b) and (c) show the estimation values and precisions for the kinetic parameters $k_a$, $k_s$ and $k_d$ as $\nu$ increases for $m=10$. For each value of $\nu$ the error bars represent the coherent and Fock states, going from left to right. Panels (d), (e) and (f) show the corresponding enhancement ratio for the Fock state for $m=10$. Panels (g), (h) and (i) show the estimation values and precisions for the kinetic parameters $k_a$, $k_s$ and $k_d$ as $N$ increases for $m=10$ and $\nu=100$. For each value of $N$ the error bars represent the coherent and Fock states, going from left to right. Panels (j), (k) and (l) show the corresponding enhancement ratio for the Fock state. In panels (d), (e), (f), (j), (k) and (l) the dotted lines are a guide representing the enhancement expected from the ratio $R_M$ at the mid-point of the sensorgram for the Fock state.}
\label{singlemodeKausaite} 
\end{figure*}

\begin{figure*}[t]
\centering
\includegraphics[width=18cm]{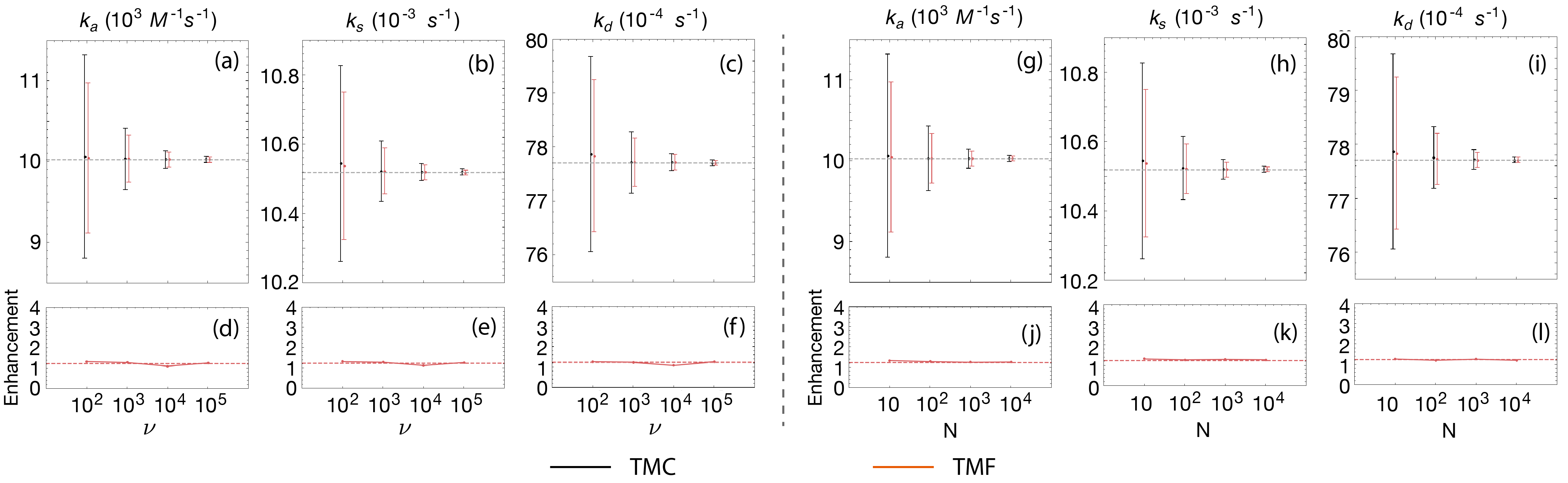}
\caption{Single-mode sensing using Fock (TMF) and coherent (TMC) states (loss, $\eta_a=0.8$ and $\eta_b=0$). Panels (a), (b) and (c) show the estimation values and precisions for the kinetic parameters $k_a$, $k_s$ and $k_d$ as $\nu$ increases for $m=10$. For each value of $\nu$ the error bars represent the coherent and Fock states, going from left to right. Panels (d), (e) and (f) show the corresponding enhancement ratio for the Fock state for $m=10$. Panels (g), (h) and (i) show the estimation values and precisions for the kinetic parameters $k_a$, $k_s$ and $k_d$ as $N$ increases for $m=10$ and $\nu=100$. For each value of $N$ the error bars represent the coherent and Fock states, going from left to right. Panels (j), (k) and (l) show the corresponding enhancement ratio for the Fock state. In panels (d), (e), (f), (j), (k) and (l) the dotted lines are a guide representing the enhancement expected from the ratio $R_M$ at the mid-point of the sensorgram for the Fock state.}
\label{singlemodeKausaiteloss} 
\end{figure*}

We now investigate the impact of loss in this optimized scenario. As the TMF state outperforms the other states, we focus on it and compare it with the TMC state. In Fig.~\ref{optimizedtwomodeKausaiteloss}(a)-(c) we show the estimation value and estimation precision for the kinetic parameters $k_a$, $k_s$ and $k_d$ for increasing $\nu$, with $\eta_a=0.8$ and $\eta_b=0.8T_{mid}$. We have used $N=10$ for the photon number and $m=10$. As in the standard two-mode sensing scenario, one can see that even in the presence of moderate loss the TMF state provides the best estimation in the kinetic parameters for any $\nu$. The enhancement is shown in Fig.~\ref{optimizedtwomodeKausaiteloss}(d)-(f) for $k_a$, $k_s$ and $k_d$ as $\nu$ increases. The enhancements are roughly in line with that expected from the mid-point value (dotted line) and independent of $\nu$. In Appendix~\ref{sec:midpointenhance} we show how the enhancement changes about the mid-point of the sensorgram for each of the states when there is loss.

In Fig.~\ref{optimizedtwomodeKausaiteloss}(g)-(i) we show the dependence of the estimation value and precision on the photon number for a fixed $\nu$. We have set $\nu=100$ and varied $N$ from 10 to 10,000. As $N$ increases the TMF state again provides the best estimation in the kinetic parameters for any $N$. In Fig.~\ref{optimizedtwomodeKausaiteloss}(j)-(l) we show the corresponding enhancement behavior.

\subsection{Single-mode sensing}

In a final scenario, we reduce the two-mode sensing model to a single-mode model by effectively removing the reference mode $b$ by setting $\eta_b = 0$. This means that there will be no transmittance in that mode and the intensity-difference measurement, $\langle \hat{M} \rangle$, becomes an intensity measurement of the signal mode. This scenario may be more feasible in an experiment. Indeed, several experiments have already demonstrated a single-mode scenario for quantum plasmonic sensing with $N=1$ Fock states~\cite{Lee2018,Peng2020,Zhao2020}. In the case of a parameter extracted from a static transmittance, the Fock state is known to be the optimal state~\cite{Nair2018,Tame19,Lee2021} and therefore we focus on using it for the case here involving parameters estimated from a dynamic transmittance.

In Fig.~\ref{singlemodeKausaite}(a)-(c) we show the estimation value and estimation precision for the kinetic parameters $k_a$, $k_s$ and $k_d$ for increasing $\nu$, with $N=10$ and $m=10$. One can see that the Fock state (TMF state with $\eta_b=0$) provides the best estimation in the kinetic parameters for any $\nu$ when compared to a coherent state in the signal mode with matched mean photon number (TMC state with $\eta_b=0$). The corresponding enhancement is shown in Fig.~\ref{singlemodeKausaite}(d)-(f) for $k_a$, $k_s$ and $k_d$ as $\nu$ increases. The enhancements are all similar and roughly in line with that expected from the mid-point value $R_M$ (dotted line) and independent of $\nu$. In Appendix~\ref{sec:midpointenhance}, we have not included the behavior of the enhancement around the mid-point in this scenario, as the Fock state enhancement in the single-mode scenario is the same as that of the TMF state in the optimized two-mode scenario for any value of $T$~\cite{Tame19}.

The dependence of the estimation value and precision on the photon number for a fixed $\nu$ is shown in Fig.~\ref{singlemodeKausaite}(g), (h) and (i) for $k_a$, $k_s$ and $k_d$, respectively. In this case, we set $\nu=100$ and vary $N$ from 10 to 10,000. One can see that as $N$ increases the Fock state provides the best estimation in the kinetic parameters for any $N$. Again, in Appendix~\ref{sec:midpointenhance}, we have not included the behavior of the enhancement around the mid-point due to its equivalence to that of the optimized two-mode scenario.

In Fig.~\ref{singlemodeKausaiteloss}(a)-(c) we consider loss in the signal mode and show the estimation value and estimation precision for the kinetic parameters $k_a$, $k_s$ and $k_d$ for increasing $\nu$, with $\eta_a=0.8$ and $\eta_b=0$. We have used $N=10$ for the photon number and $m=10$. As before, one can see that even in the presence of moderate loss the Fock state provides the best estimation in the kinetic parameters for any $\nu$. The enhancement is shown in Fig.~\ref{singlemodeKausaiteloss}(d)-(f) for $k_a$, $k_s$ and $k_d$ as $\nu$ increases. The enhancements are roughly in line with that expected from the mid-point value (dotted line) and independent of $\nu$. 

In Fig.~\ref{singlemodeKausaiteloss}(g)-(i) we show the dependence of the photon number for a fixed $\nu$. We have set $\nu=100$ and varied $N$ from 10 to 10,000. As $N$ increases the Fock state again provides the best estimation in the kinetic parameters for any $N$. In Fig.~\ref{singlemodeKausaiteloss}(j)-(l) we show the corresponding enhancement behavior.

It is interesting to note that in static quantum plasmonic sensing with a single mode, the Fock state provides an enhancement in the estimation precision for any value of $\eta_a$ (see Refs.~\cite{Lee2021,Tame19}). In this case, the enhancement tends to unity as $\eta_a$ goes to zero. While our dynamic results are limited to the case of $\eta_a=1$ and $\eta_a=0.8$, we have shown that the enhancement carries over well from the static to the dynamic case and that it is mainly determined by the enhancement around the mid-point value. It is therefore likely that the static enhancement carries over to the dynamic case for any value of $\eta_a$. Further work in this direction would be needed to confirm such behavior.

\begin{figure}[t]
\centering
\includegraphics[width=8cm]{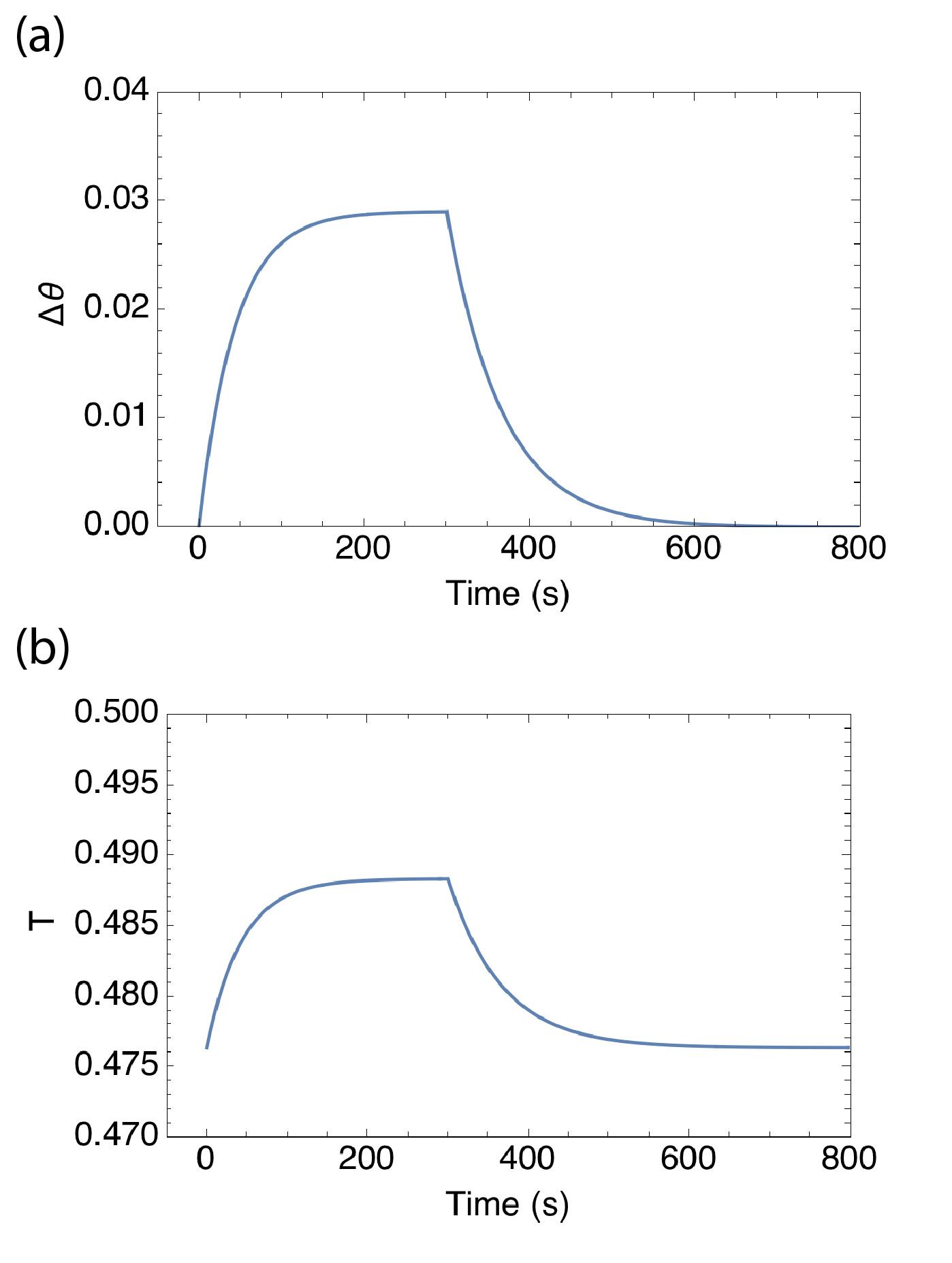}
\caption{Sensorgrams from the experiment by Lahiri {\it et al.}~\cite{Lahiri99} which investigates CA interacting with benzenesulfonamide. (a) Angular sensorgram, $\Delta\theta(t)$, where the full sensorgram is $\theta(t)=\theta(0)+\Delta\theta(t)$ with $\theta(0)=66.796$ degrees. (b) Reconstructed transmittance sensorgram, $T(t)$, with $\theta_{in}=66.21$ degrees set.}
\label{angtranssensorgram2} 
\end{figure}

\section{Small sensorgram deviation}

We now study a second interaction, that of carbonic anhydrase (CA) interacting with the inhibitor benzenesulfonamide, which is well documented in the work of Lahiri {\it et al.}~\cite{Lahiri99}. This second interaction process was chosen as it produces a small sensor response, or deviation, due to the small change in the refractive index during the interaction dynamics. It therefore gives information about the performance of quantum techniques for measuring kinetic parameters in pathogen-inhibitor interactions relevant to drug development. In this case, usually a much larger pathogen is immobilized as the receptor and the smaller inhibitor molecule is added as the ligand in the analyte~\cite{Shankaran2007,Homola2008}. Due to the small size of the ligand molecule, the resulting refractive index change is much smaller.

\subsection{Transmittance sensorgram}

In the experiment reported by Lahiri {\it et al.}~\cite{Lahiri99}, the CA is immobilized on a gold surface using a self-assembled monolayer following a similar method to that used in Kausaite {\it et al.}~\cite{Kausaite07}, in the previous section. The real-time SPR curve is measured for the CA interacting with benzenesulfonamide in an analyte using a BIAcore 1000 commercial SPR sensor developed by BIAcore~\cite{Biacore}. From the sensorgram obtained they extracted the following kinetic parameters, ${k_a = 3.8 \times 10^{-3}~{\rm M}^{-1} {\rm s}^{-1}}$, ${k_d = 15 \times 10^{-3}~{\rm s}^{-1}}$ and ${L_0 = 2.1~{\rm M}}$. However, as before, in the experiment angular interrogation is used and therefore the sensorgram obtained is angle dependent. A transformation is required to go from the angular sensorgram to the corresponding transmittance sensorgram that can be used to compare the different quantum states.

We follow the method outlined in the previous section in order to obtain the transmittance sensorgram. The angular sensorgram $\Delta \theta (t)$ is shown in Fig.~\ref{angtranssensorgram2}(a) and has the following parameters: ${A}_{\infty}=0.0291$ and $\tau=300$ (see Appendix~\ref{sec:Tsensor} for details). We use this model to find the sensorgram for intensity interrogation, $T(t)$. At $t=0$, we set $n^2_a(0)$ to be equal to that of the buffer solution used in the experiment, which for PBS is 1.3385~\cite{Dieguez2009}. Setting the frequency $\omega$ corresponding to the wavelength of the laser used ($\lambda=760$ nm), with $n_p=1.523$~\cite{Lahiri99} and using $\epsilon_m=-20.913+i1.2923$~\cite{Johnson1972}, we obtain from Eq.~\eqref{thetadep} in Appendix \ref{sec:Tsensor} the angle $\theta(0)=66.796$ degrees. With the time dependence of $n_a(t)$ known from Eq.~\eqref{reftime}, we then use it in Eq.~\eqref{eqn:r} to obtain $T(t)=|r_{spp}(t)|^2$, which is shown in Fig.~\ref{angtranssensorgram2}(b). In this plot we have set $\theta_{in}=66.21$ degrees, which corresponds to an angle below $\theta(0)$ where we are operating close to the inflection point for the transmittance curve, as shown in the inset of Fig.~\ref{fig1}(a). We have set the thickness of the gold as $d=38$~nm, as used in the experiment~\cite{Lahiri99}.

Due to the small change in the transmittance $T$ during the interaction, the response of the sensor can be assumed to be linear. Thus, the reconstructed transmittance sensorgram does not require any calibration. The kinetic parameters obtained from the ideal $T(t)$ sensorgram shown in Fig.~\ref{angtranssensorgram2}(b) are ${k_s = 22.98 \times 10^{-3}~{\rm s}^{-1}}$, ${k_d = 15 \times 10^{-3}~{\rm s}^{-1}}$ and ${k_a = 3.8 \times 10^{-3}~{\rm M}^{-1} {\rm s}^{-1}}$, where we have used ${L_0 = 2.1~{\rm M}}$ to extract out $k_a$ from the parameters $k_s$ and $k_d$, as before. Due to the linear transmittance response of the sensor, these parameters match exactly those obtained from the angular sensorgram.

We now consider using the classical state (TMC state) and compare the estimate and precision of the kinetic parameters obtained with it to those obtained using the quantum states. As in the previous section, we simulate the measurement process and noise according to the Monte Carlo simulation method described in Section II C. For conciseness we focus on only one of the kinetic parameters, the association constant $k_a$. This kinetic parameter depends on the other two parameters $k_s$ and $k_d$ extracted from the fits using the relation ${k_a}=({k_s}-k_d)/ [{L_0}]$ and it therefore gives an idea of the overall enhancement in the estimation of the kinetic parameters. However, the estimation precisions of the individual parameters $k_s$ and $k_d$ were checked and found to individually follow the same behavior as $k_a$. In addition, for the low photon number of $N=10$, the value of the set size $\nu$ for each instance of time in the sensorgram had to be increased considerably as the noise in the measurement signal was too large otherwise (low signal-to-noise ratio) and a fit could not be obtained. We have chosen to fix $\nu=10^5$, as values above this are challenging to reach in an experiment~\cite{Lee2018}. Here, we focus on the dependence of the estimation value and its precision as the photon number increases. We have also chosen to focus on the comparison between the TMC and TMF states, as it is clear that the TMSV and TMSD states only improve on the TMC state in the two-mode optimized scenario, and in this case the enhancement in precision is similar to that of the TMF state in both the static case~\cite{Tame19,Lee2021} and dynamic case for large deviation (see previous section).

\subsection{Standard two-mode sensing}

\begin{figure}[t]
\centering
\includegraphics[width=8.5cm]{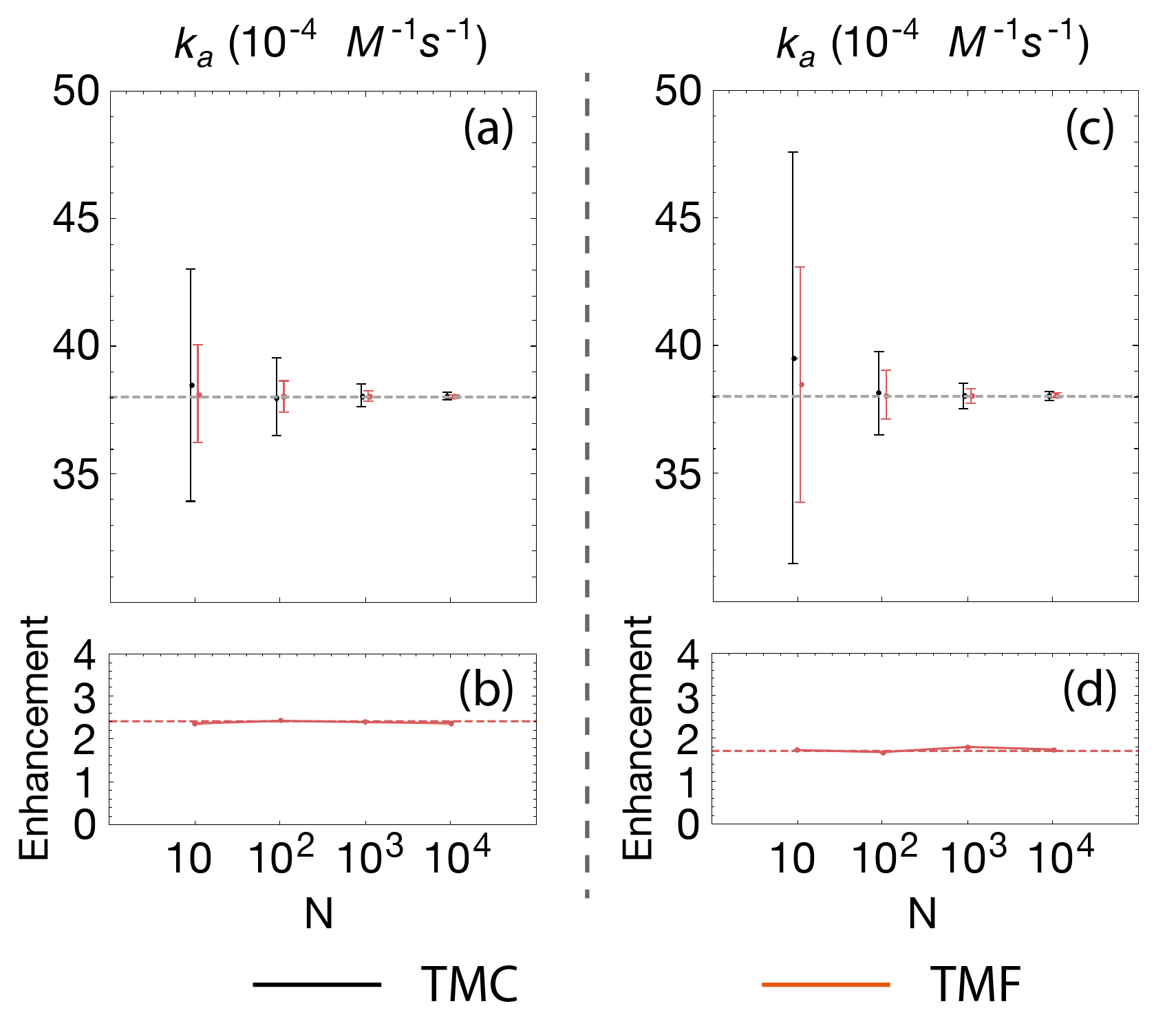}
\caption{Photon number dependence in standard two-mode sensing using TMC and TMF states for a small sensorgram deviation. Panels (a) and (c) show the estimation values and precisions for the kinetic parameter $k_a$ as $N$ increases for $m=10$ and $\nu=10^5$, where (a) corresponds to no loss ($\eta_a=\eta_b=1$) and (c) corresponds to loss ($\eta_a=\eta_b=0.8$). For each value of $N$ the error bars represent the TMC and TMF states, going from left to right. Panels (b) and (d) show the corresponding enhancement ratio for the TMF state, where (b) corresponds to no loss ($\eta_a=\eta_b=1$) and (d) corresponds to loss ($\eta_a=\eta_b=0.8$). The dotted lines are a guide representing the enhancement expected from the ratio $R_M$ at the mid-point of the sensorgram for the TMF state.}
\label{standardtwomodeLahiri} 
\end{figure}

In this first scenario, as before, we consider the general sensing model shown in Fig.~\ref{fig1}(c), where we set the loss in either mode to be the same. To start with, we take the ideal case of $\eta_a = \eta_b = 1$. In Fig.~\ref{standardtwomodeLahiri}(a) we show the estimation value and estimation precision for the kinetic parameter $k_a$ as we vary $N$ from 10 to 10,000. As before, the estimation precision shown for each state physically corresponds to that of a fixed set of $m=10$ sensorgrams, each of which has $\nu$ states probed at a given instance of time, with a step-size between instances of time of 5s, as described in more detail in Section II C. In the present scenario, the sensorgram is 1000 seconds in duration and so there are 200 points in total, each point having $\nu$ probe states measured. The step-size for the points was chosen so that there was a fine enough mesh for the fit to return the exact values in the ideal case when there is no noise.

In Fig.~\ref{standardtwomodeLahiri}(a), one can see that as $N$ increases the TMF state provides the best estimation in the kinetic parameters for any $N$. The enhancement plot is shown in Fig.~\ref{standardtwomodeLahiri}(b). In this standard two-mode sensing scenario, the TMF state is clearly the state that offers the best estimation precision. In Fig.~\ref{standardtwomodeLahiri}(c) we show the estimation value and estimation precision for the kinetic parameter $k_a$ for increasing photon number $N$, with $\eta_a=\eta_b=0.8$. One can see that even in the presence of moderate loss the TMF state provides the best estimation in the kinetic parameters.

\subsection{Optimized two-mode sensing}

\begin{figure}[t]
\centering
\includegraphics[width=8.5cm]{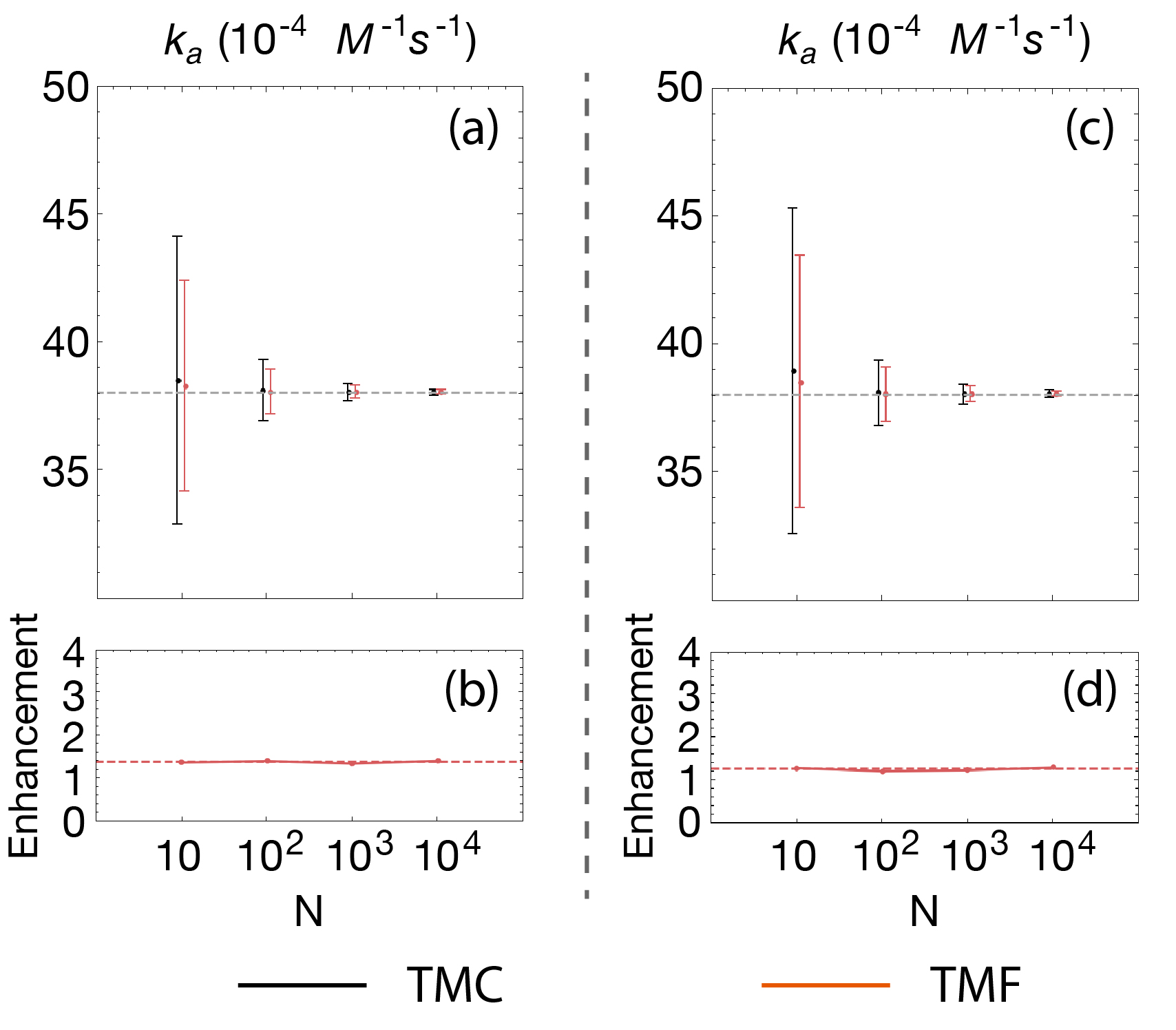}
\caption{Photon number dependence in optimized two-mode sensing using TMC and TMF states for a small sensorgram deviation. Panels (a) and (c) show the estimation values and precisions for the kinetic parameter $k_a$ as $N$ increases for $m=10$ and $\nu=10^5$, where (a) corresponds to no loss ($\eta_a=1$ and $\eta_b=T_{mid}$) and (c) corresponds to loss ($\eta_a=0.8$ and $\eta_b=0.8T_{mid}$). For each value of $N$ the error bars represent the TMC and TMF states, going from left to right. Panels (b) and (d) show the corresponding enhancement ratio for the TMF state, where (b) corresponds to no loss ($\eta_a=1$ and $\eta_b=T_{mid}$) and (d) corresponds to loss ($\eta_a=0.8$ and $\eta_b=0.8T_{mid}$). The dotted lines are a guide representing the enhancement expected from the ratio $R_M$ at the mid-point of the sensorgram for the TMF state.}
\label{optimzedtwomodeLahiri} 
\end{figure}

In this second scenario, we set $\eta_b = \eta_aT_{mid}$, with $T_{mid}=0.4824$. In Fig.~\ref{optimzedtwomodeLahiri}(a) we show the estimation value and estimation precision for the kinetic parameter $k_a$ as we vary $N$ from 10 to 10,000 for the ideal case of $\eta_a =1$ and $\eta_b = T_{mid}$. One can see that as $N$ increases the TMF state provides the best estimation in the kinetic parameters for any $N$. The enhancement plot is shown in Fig.~\ref{optimzedtwomodeLahiri}(b). In this optimized two-mode sensing scenario, the TMF state is clearly the state that offers the best estimation precision. In Fig.~\ref{optimzedtwomodeLahiri}(c) we show the estimation value and estimation precision for the kinetic parameter $k_a$ for increasing photon number $N$, with $\eta_a=0.8$ and $\eta_b=0.8T_{mid}$. One can see that even in the presence of moderate loss the TMF state again provides the best estimation in the kinetic parameters.

\subsection{Single-mode sensing}

\begin{figure}[t]
\centering
\includegraphics[width=8.5cm]{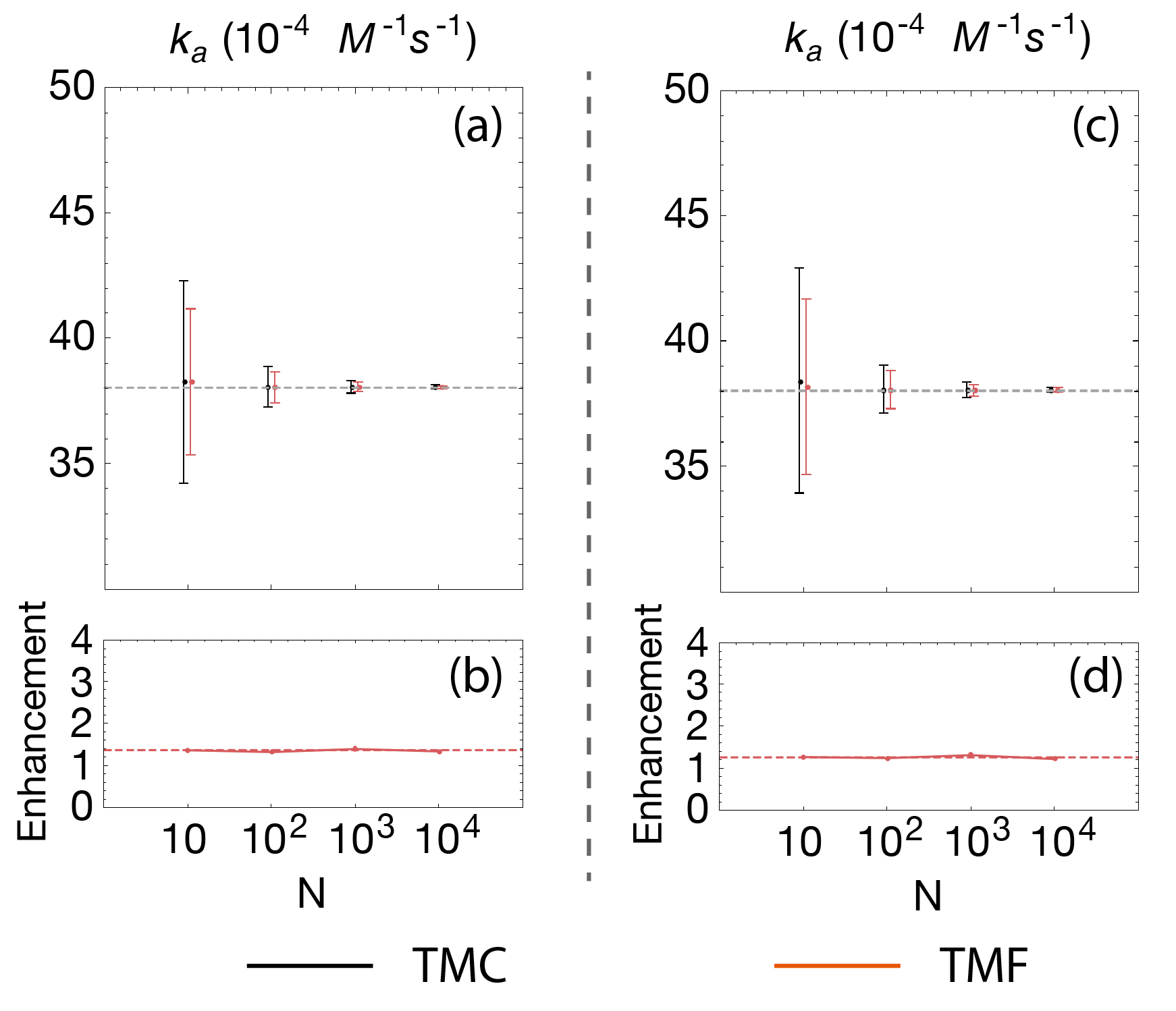}
\caption{Photon number dependence in single-mode sensing using coherent (TMC) and Fock (TMF) states for a small sensorgram deviation. Panels (a) and (c) show the estimation values and precisions for the kinetic parameter $k_a$ as $N$ increases for $m=10$ and $\nu=10^5$, where (a) corresponds to no loss ($\eta_a=1$ and $\eta_b=0$) and (c) corresponds to loss ($\eta_a=0.8$ and $\eta_b=0$). For each value of $N$ the error bars represent the coherent and Fock states, going from left to right. Panels (b) and (d) show the corresponding enhancement ratio for the Fock state, where (b) corresponds to no loss ($\eta_a=1$ and $\eta_b=0$) and (d) corresponds to loss ($\eta_a=0.8$ and $\eta_b=0$). The dotted lines are a guide representing the enhancement expected from ratio $R_M$ at the mid-point of the sensorgram for the Fock state.}
\label{singlemodeLahiri} 
\end{figure}

In the final scenario, we reduce the two-mode sensing model to a single-mode model by effectively removing the reference mode $b$ by setting $\eta_b = 0$. As mentioned before, this means that there is no transmittance in that mode and the intensity difference measurement effectively becomes an intensity measurement of the signal mode. In Fig.~\ref{singlemodeLahiri}(a) we show the estimation value and estimation precision for the kinetic parameter $k_a$ as we vary $N$ from 10 to 10,000 for the ideal case of $\eta_a =1$ and $\eta_b = 0$. One can see that as $N$ increases the TMF state provides the best estimation in the kinetic parameters for any $N$. The enhancement plot is shown in Fig.~\ref{singlemodeLahiri}(b). In this single-mode sensing scenario, the TMF state is again the state that offers the best estimation precision. In Fig.~\ref{singlemodeLahiri}(c) we show the estimation value and estimation precision for the kinetic parameter $k_a$ for increasing photon number $N$, with $\eta_a=0.8$ and $\eta_b=0$. Thus, even in the presence of moderate loss the TMF state provides the best estimation in the kinetic parameters.

\subsection{Threshold Intensity}

In the case of a small sensorgram deviation it was pointed out that due to the signal-to-noise ratio a large value of $\nu$ is needed to be able to make a fit and extract out the kinetic parameters for the classical TMC state and TMF state, regardless of the sensing scenario. This is interesting because it implies that for a fixed mean photon number $N$ in the probe mode, a relatively high intensity (high probe state rate or large $\nu$) is needed for studying interactions that result in a small sensorgram deviation. If the biosystems (receptors and ligands) being studied are fragile to such a high intensity it would potentially add extra noise to the measurements and make them less precise~\cite{Bartczak2011,Robinson2014,Taylor2016}, an effect recently observed for $3\mu$m polystyrene beads~\cite{Casacio2021}. The sensor itself may also impart additional noise or nonlinear behavior at a high intensity~\cite{Piliarik2009,Kaya2013}. 

Regardless of the size of the sensorgram deviation, in the case where the additional noise mentioned above is introduced, it is favourable to operate at a lower intensity. Let $I_t=N r_t$ be a threshold intensity above which the biosystem or sensor does not function correctly and assume a fixed interaction region of the sensor that a state probes (spatial area of the signal mode profile). Here, $r_t$ is the threshold rate of states used for probing, each state having mean photon number $N$ in the signal mode. Assuming a sensorgram represents an ergodic process over at most 1s, this means that at the threshold we have $m \nu \leq r_t$, {\it i.e.}, for a fixed $m$, $r_t$ sets an upper bound on $\nu$. At the threshold, to improve the signal-to-noise ratio one can increase $N$, as in general the ratio scales as $({N \nu})^{1/2}$ for all the states considered (see Appendix~\ref{sec:noise}). However, $r_t$ will need to decrease and therefore $\nu$ decreases, with the ratio remaining unchanged. The parameters $N$ and $\nu$ are interchangeable in this sense and conversely one can also try to improve the signal-to-noise ratio by increasing $\nu$ via increasing $r_t$ at the expense of decreasing $N$. A higher signal-to-noise ratio would lead to an enhancement in the estimation precision of the kinetic parameters, but it appears that this is not possible at the threshold intensity by changing only $N$ or $\nu$.

In this work we have shown that for a given $N$, a quantum state, {\it e.g.}, the TMF state, can offer the same estimation precision as the classical TMC state at a higher $N$ (for a fixed $\nu$). We have also shown that for a given $\nu$, a quantum state can offer the same precision as the classical TMC state at a higher $\nu$ (for a fixed $N$). Thus, in general the precision of the classical state can be obtained with a quantum state using a lower intensity. This means that one can achieve a better precision than the classical TMC state at the threshold intensity -- the point at which the maximum precision is achieved for any state.

\section{Summary and Outlook}

In this work we studied theoretically the measurement of the kinetic parameters of two interaction processes using a plasmonic resonance sensor and quantum states of light. The first interaction studied was BSA interacting with IgG1 (anti-BSA), which was chosen as it produces a large sensor response. The second interaction studied was carbonic anhydrase interacting with benzenesulfonamide, which was chosen as it produces a small sensor response and therefore gives information about the performance of quantum techniques for measuring kinetic parameters of interactions relevant to drug development. 

We started by introducing the physical model for plasmonic sensing and provided details of the sensor setup, its response to a dynamically changing environment, the general model for interaction kinetics, and the various quantum states and measurements. For the interaction processes we considered the classical TMC state and various quantum states: the TMF, TMSV and TMSD states. We also described the simulation method we used to model noise in the sensor's measurement signal for these states.

Our results show that the enhancement in the estimation precision offered by quantum states translates well from static to kinetic parameters, such as the association and dissociation binding constants of the interactions. In the case of a large sensorgram deviation, a wide range of $\nu$, $m$ and $N$ values could be considered and an enhancement was shown for the various quantum states in three main scenarios: standard, optimized and single-mode. On the other hand, in the case of a small sensorgram deviation, a large value of $\nu$ was needed to be able to make a fit and extract out the kinetic parameters for the classical TMC state and TMF state. This lead to a discussion about intensity and additional intensity-based noise due to the fragility of biosystems and the sensor itself. It was mentioned that the classical precision in kinetic parameters can be obtained with a quantum state using a lower intensity and that therefore one can achieve a better precision than the classical TMC state at the threshold intensity -- the point at which the maximum precision is achieved for any state before additional noise is introduced. 

While the enhancement in the precision found using quantum states is small at around 1-3 times that of the classical case, even such a small improvement in the estimation precision could make a big difference in accurately determining the kinetic parameters when operating close to the intensity and noise limits of a sensor.

The insights and analysis provided in this study may help in designing quantum plasmonic sensors that enable more precise kinetics research. Future work might look at extending the analysis given here to angular interrogation and other interesting kinetic interactions in pharmaceutical research.

{\it Acknowledgements.---} This research was supported by the South African National Research Foundation, the National Laser Centre and the South African Research Chair Initiative of the Department of Science and Innovation and National Research Foundation. C. L. is supported by a KIAS Individual Grant (QP081101) via the Quantum Universe Center at Korea Institute for Advanced Study and Korea Research Institute of Standards and Science (KRISS–GP2022-0012). G. E. M. M. and H. G. K. thank UKZN for financial and administrative support.


\appendix
\onecolumngrid
\newpage

\section{Measurement noise of the quantum states \label{sec:noise}}

In this appendix we give the formulas for the noise in the measurement of the different quantum states in the standard two-mode scenario, {\it i.e.}, $\Delta M$. For the TMC state we have~\cite{Lee2021}
\be
\Delta M_{TMC}=(\eta_a T N_a+\eta_b N_b)^{1/2},
\label{DeltaTMC}
\ee
where $N_a=N_b=N$ in the symmetric (balanced) case for the enhancement calculation, $R_M=\Delta M_C/\Delta M_Q$, {\it i.e.}, when the classical TMC state ($C=TMC$) is compared with a quantum state $Q$ with equal number of photons in the signal and reference modes ($Q=TMF$ and $TMSV$). For the TMF and TMSV states we have~\cite{Lee2017}
\bqa
\Delta M_{TMF}&=& N^{1/2} (\eta_a T(1-\eta_aT)+\eta_b(1-\eta_b))^{1/2}, \label{DeltaTMF} \\
\Delta M_{TMSV}&=& N^{1/2} ((T \eta_a-\eta_b)^2N+\eta_b+T\eta_a(1-2\eta_b))^{1/2}. \label{DeltaTMSV}
\eqa
For the TMSD state we have
\bqa
&&\Delta M_{TMSD}=[2T^2\eta_a^2G(G-1) |\alpha|^2+T^2\eta_a^2(G-1)^2 + T\eta_a(G|\alpha|^2+(G-1))+2\eta_b^2(G-1)^2|\alpha|^2  \nonumber \\
&&\qquad \qquad \qquad +\eta_b^2(G-1)^2+\eta_b((G-1)|\alpha|^2+G-1)-4T\eta_a\eta_bG(G-1)|\alpha|^2-2T\eta_a\eta_bG(G-1)]^{1/2}. \label{DeltaTMSD}
\eqa
In the above we have used $G=\cosh^2 r$. The formula can be put in terms of $N$ using the relation $N=G|\alpha|^2+(G-1)$ and is obtained using the relation
\be
\Delta M=[\langle \hat{M}^2 \rangle-\langle \hat{M} \rangle^2]^{1/2}=[\Delta N_a^2+\Delta N_b^2 - 2 (\langle \hat{N}_a \hat{N}_b\rangle-\langle \hat{N}_a \rangle \langle \hat{N}_b \rangle)]^{1/2}.
\label{DeltaTMSD2}
\ee
The expectation values of Eq.~\eqref{DeltaTMSD2} can be calculated in the Heisenberg picture, where the initial state before squeezing, $\ket{\alpha}_a\ket{0}_b$, is used together with the evolution of the operators as~\cite{Tame19,Lee2021}
\bqa
\hat{a}&=&\sqrt{GT\eta_a} \hat{a}+\sqrt{(G-1)T\eta_a}\hat{b}^\dag+\sqrt{1-T}\hat{c}+\sqrt{T(1-\eta_a)}\hat{d}, \\
\hat{b}&=&\sqrt{G\eta_b} \hat{b}+\sqrt{(G-1)\eta_b}\hat{a}^\dag+\sqrt{1-\eta_b}\hat{e}.
\eqa
The operators $\hat{c}$, $\hat{d}$ and $\hat{e}$ represent noise operators for modes that are initially in the vacuum state. One finds the following terms
\bqa
\Delta N_a^2 &=& T^2 \eta_a^2(G-1)[(G-1)+2G|\alpha|^2]+T\eta_a[(G-1)+G|\alpha^2|], \\
\Delta N_b^2 &=& (G-1)^2\eta_b^2(2|\alpha|^2+1)+(G-1)\eta_b(|\alpha|^2+1), \\
\langle \hat{N}_a \hat{N}_b\rangle&=& T\eta_a\eta_b[G(G-1)(|\alpha|^4+2|\alpha|^2)+G(G-1)(|\alpha|^2+1)+(G-1)^2(|\alpha|^2+1)], \\
\langle \hat{N}_a \rangle&=& T\eta_a(G|\alpha|^2+(G-1)), \\
\langle \hat{N}_b \rangle&=& \eta_b(G-1)(|\alpha|^2+1).
\eqa
In the limit of $|\alpha|^2 \gg1$ we have from Eq.~\eqref{DeltaTMSD},
\bqa
&&\Delta M_{TMSD}=|\alpha|[T\eta_aG+\eta_b(G-1)+2(G-1)(G(T\eta_a-\eta_b)^2-\eta_b^2)]^{1/2},
\eqa
which leads to the formula in Refs.~\cite{Tame19,Lee2017} for $\Delta M_{TMSD}^2/\Delta M_{TMC}^2$ when $N_a$ and $N_b$ in $\Delta M_{TMC}$ (see Eq.~\eqref{DeltaTMC}) are set to match the initial values for the TMSD state.

In the optimized two-mode scenario we then set $\eta_b=\eta_aT_{mid}$ in Eqs.~\eqref{DeltaTMC}, \eqref{DeltaTMF}, \eqref{DeltaTMSV} and \eqref{DeltaTMSD}. In the single-mode scenario we set $\eta_b=0$.

\section{Interaction kinetics \label{sec:intkin}}

In order to model the dynamics of a given interaction, the receptor-ligand complex concentration $[C]$ at any time in the system should be found. The increasing rate of $[C]$ at an instance of time in the system is ${k_a}[{L}][{R}]$. Where $[L]$ and $[R]$ are the individual ligand and receptor concentrations at that time in units of ${\rm M}$ (molarity, or moles per liter - mol/l), respectively, and ${k_a}$ is the association constant measured in ${\rm M}^{-1} {\rm s^{-1}}$ (per molarity per second), which is determined by collision rates involved in the interactions between ligand and receptor molecules. At ${t}=0$, the ligand concentration can be taken as $[{L_0}]$ and the receptor concentration as $[{R_0}]$. As the concentration of the receptor-ligand complex increases, the concentration of the individual ligand and receptor molecules decreases. At ${t \geq 0}$, we have the concentration of ligands $[{L}]=[{L_0}]-[{C}]$ and the concentration of receptors $[{R}]=[{R_0}]-[{C}]$. One can also look at backward reactions, where the decreasing rate of the complex concentration is $k_d[C]$, where ${k_d}$ is the dissociation constant measured in ${\rm s^{-1}}$. At equilibrium, the increasing rate and decreasing rate of the complex should be the same, {\it i.e.}, ${k_a [L]~[R]= k_d [C]}$. From this equality, one can obtain the dissociation equilibrium constant of the ligand-receptor interaction given by ${K_D}=\frac{{k_d}}{{k_a}}=\frac{[{L}]~[{R}]}{{[C]}}$, which is in units of ${\rm M}$. The reciprocal of ${K_D}$, given by ${K_A}=1/K_D$, is called the affinity of the ligand-receptor interaction, with units of ${\rm M^{-1}}$. 

In the association phase, consider the initial concentrations of the two `reactants' are $[{L_0}]$ and $[{R_0}]$ for ligands and receptors, respectively, above the gold surface in the flow cell of the plasmonic sensor shown in Fig.~\ref{fig1}(a). The complex concentration then changes with time before the reactions reach equilibrium. The evolution of $[C]$ is given by
\bqa
\frac{{d}[{C}]}{{d t}}&=&{k_a}[{R}][{L}]-{k_d} [{C}] \nonumber \\
&=&{k_a}([{R_0}]-[{C}])([{L_0}]-[{C}])-{k_d} [{C}].
\label{conc}
\eqa
The reaction of the ligand-receptor is therefore a second-order process. However, one can solve the equation as a pseudo-first-order approximation. This is valid when supplying the ligand concentration to the flow cell in great excess to the receptor concentration, which implies that $[{L_0}] \gg [{R_0}] $. Hence the amount of ligand used in the binding interactions is negligible compared to the initial ligand concentration, {\it i.e.}, $[{L_0}]-[{C}] \approx [{L_0}] $. From this, we then have from Eq.~\eqref{conc} 
\begin{equation}
\frac{{d}[{C}]}{{d t}}={k_a}([{R_0}]-[{C}])[{L_0}]-{k_d} [{C}].
\end{equation}
The solution to this first-order equation for an initial complex concentration of $[C_0]=0$ is given as
\begin{equation}
[{C}] = \frac{[{L_0}][{R_0}]}{[{L_0}]+ {k_d/k_a}}(1-{e^{({k_a} [{L_0}]+ {k_d}){t} }}).
\end{equation}
It is clear that the complex concentration increases exponentially with time and for $t \gg 0$ reaches a steady state value. This is the steady state phase of the interaction kinetics.

We now consider the dissociation phase. At time ${t} = \tau$, the complex concentration in the chamber has increased to $[{C_{\tau}}]$. At this point the flow cell is then washed by a buffer solution, {\it e.g.}, water in an elution process, which means that once a ligand unbinds from a receptor the possibility of it binding to the same or another receptor is negligible. The possibility of another ligand binding to the receptor is also negligible as the background ligand concentration in the flow cell is effectively zero. With the initial conditions of $[{C_{\tau}}]$ and $[L_0]\approx 0$ in Eq.~\eqref{conc}, the solution is given by
\begin{equation}
\frac{{d}[{C}]}{{d t}}=-{k_d} [{C}].
\end{equation}
From this solution we see the complex concentration decreases exponentially with time from the start of the elution process at ${t} = \tau$.

The concentration of the complex $[C]$ and the transmittance of the sensor $T$ are linked by the refractive index, $n_a=\sqrt{\epsilon_a}$, of the region above the gold surface, whose change is induced by the sequence of analytes being passed over the flow cell: (i) buffer and ligands (association and steady state) and (ii) buffer only (dissociation). The refractive index change can be understood as a change in the dipole moments of the immobilized receptors as they are converted into complexes and then unconverted~\cite{Xiao19}. For a fixed incidence angle of light, an increase in the complex concentration $[C]$ therefore increases the value of $\epsilon_a$ and thus $T$, as shown in the inset of Fig.~\ref{fig1}(a). In the ideal case, when there is a linear relation between $[C]$ and $T$ we can write~\cite{Xiao19} 
\begin{equation}
T(t) =
\begin{cases}
                                  {T}_{\infty}(1-{e}^{-{k_s t}})& \text{$0 \leq t < \tau $} \\
                                   {T}_{\tau} {e}^{-{k_d (t-\tau)}} & \text{$t\geq \tau$}, \\
\end{cases}
\label{sensorgrameq}
\end{equation}
where $T_\infty$ is a constant determined by the initial concentration of the ligands and receptors, the thickness of the receptor and ligand layers above the gold surface, and the affinity $k_A$. We then have the constant $T_\tau=T_\infty(1-e^{-k_s \tau})$. In the above, the constant ${k_s}={k_a}[{L_0}]+ {k_d}$ represents the observable rate for the association phase. Equation~\eqref{sensorgrameq} is the theoretical model for the sensor's response, which is the sensorgram that would be measured in an ideal experiment (no noise). From the measured sensorgram a nonlinear fit is then performed, {\it e.g.}, Gauss-Newton, with respect to the theoretical model in order to extract out the association and dissociation kinetic parameters. From the fit, $k_d$ and $k_s$ are obtained and with a knowledge of the initial ligand concentration $[L_0]$, $k_a$ can be found from the relation ${k_a}=({k_s}-k_d)/ [{L_0}]$. All simulations and fittings were done using Mathematica.

\section{Extraction of $T$ sensorgram \label{sec:Tsensor}}

We start with the expression $\theta({t})=\theta({0}) + \Delta\theta ({t})$, where $\theta({t})$ is the resonance angle in degrees at a given time (the angle causing $|r_{spp}|^2$ in Eq.~\eqref{eqn:r} to reach its minimum) and $\Delta\theta ({t})$ is a shift in that angle due to a change in the refractive index above the gold surface. The angular sensorgram is similar to that for the transmittance sensorgram given in Eq.~\eqref{sensorgrameq} and we have~\cite{Xiao19}
\begin{equation}
  \Delta\theta ({t}) =
\begin{cases}
                                  {A}_{\infty}(1-{e}^{-{k_s t}})& \text{$0 \leq t < \tau $}, \\
                                   {A}_{\tau} {e}^{-{k_d (t-\tau)}} & \text{$t\geq \tau$}.        \\
\end{cases}
\label{angleint}
\end{equation}
The value of $\tau$ is 1100 in the experiment and ${A}_{\infty}$ is measured to be $800 \times 10^{-3}$ degrees. With these values and the values of $k_a$, $k_d$ and $L_0$ stated above we have a complete model of the sensorgram for angular interrogation, as shown in Fig.~\ref{angtranssensorgram}(a). We now use this model to find the sensorgram for intensity interrogation, $T(t)$.

When the field in the signal mode is on resonance with the surface plasmon on the surface of the gold, causing $|r_{spp}|^2$ to reach its minimum, the following resonance condition holds for the component of the wavevector parallel to the surface, $\epsilon_p^2 k_0 \sin \theta_{in}=k_0[(\epsilon_m'\epsilon_a)/(\epsilon_m'+\epsilon_a)]^{1/2}$~\cite{Maier2007}, where $k_0=\omega/c$ and $\epsilon_m'=Re[\epsilon_m]$. On the left is the wavevector for the field in the prism and on the right is the wavevector of the SPP. For a time varying $\epsilon_a=\sqrt{n_a}$, we have that the angle $\theta_{in}$ satisfying the resonance condition gains a time dependence, giving $\theta(t)$, where we have dropped the subscript $in$ for convenience. We can therefore rewrite the resonance condition as
\begin{equation}
\theta({t})={\sin^{-1}} \left( \frac{\sqrt{{n^2_a(t) n^2_m}}}{{n_p} \sqrt{{n^2_a(t) + n^2_m}}}\right),
\label{thetadep}
\end{equation}
where $n^2_m=\epsilon_m'$. Rearranging the above equation gives 
\begin{equation}
{n_a (t)}=\left(\frac{{n^2_p n^2_m}{\sin}^2{\theta(t)}}{{n_m}^2- {n^2_p {\sin}^2{\theta(t)}}}\right)^{1/2}.
\label{reftime} 
\end{equation}
Thus, a knowledge of $\theta({t})=\theta({0}) + \Delta\theta ({t})$ provides the time dependence of the analyte refractive index. The final parameter to obtain is $\theta(0)$, as only the time dependence of $\Delta \theta (t)$ is known from the measured sensorgram. At $t=0$, we set $n^2_a(0)$ to be equal to that of the buffer solution used in the experiment, which for phosphate buffer solution (PBS) is 1.3385~\cite{Dieguez2009}. Setting the frequency $\omega$ corresponding to the wavelength of the laser used ($\lambda=670$ nm), with $n_p=1.5107$~\cite{ECOChemie} and using $\epsilon_m=-14.358+i1.0440$~\cite{Johnson1972}, we obtain from Eq.~\eqref{thetadep} the angle $\theta(0)=71.0966$ degrees. With the full time dependence of $n_a(t)$ now known from Eq.~\eqref{reftime}, we can use it in Eq.~\eqref{eqn:r} to obtain $T(t)=|r_{spp}(t)|^2$, which is shown in Fig.~\ref{angtranssensorgram}(b) as a solid line. In this plot we have set $\theta_{in}=70.1200$ degrees, which corresponds to an angle below $\theta(0)$ where we are operating close to the inflection point for the transmittance curve, as shown in the inset of Fig.~\ref{fig1}(a). We have set the thickness of the gold as $d=50$~nm, as used in the experiment~\cite{Kausaite07}.

\section{Enhancement in the precision dependence on $m$ \label{sec:enhancem}}

\begin{figure*}[t]
\centering
\includegraphics[width=16cm]{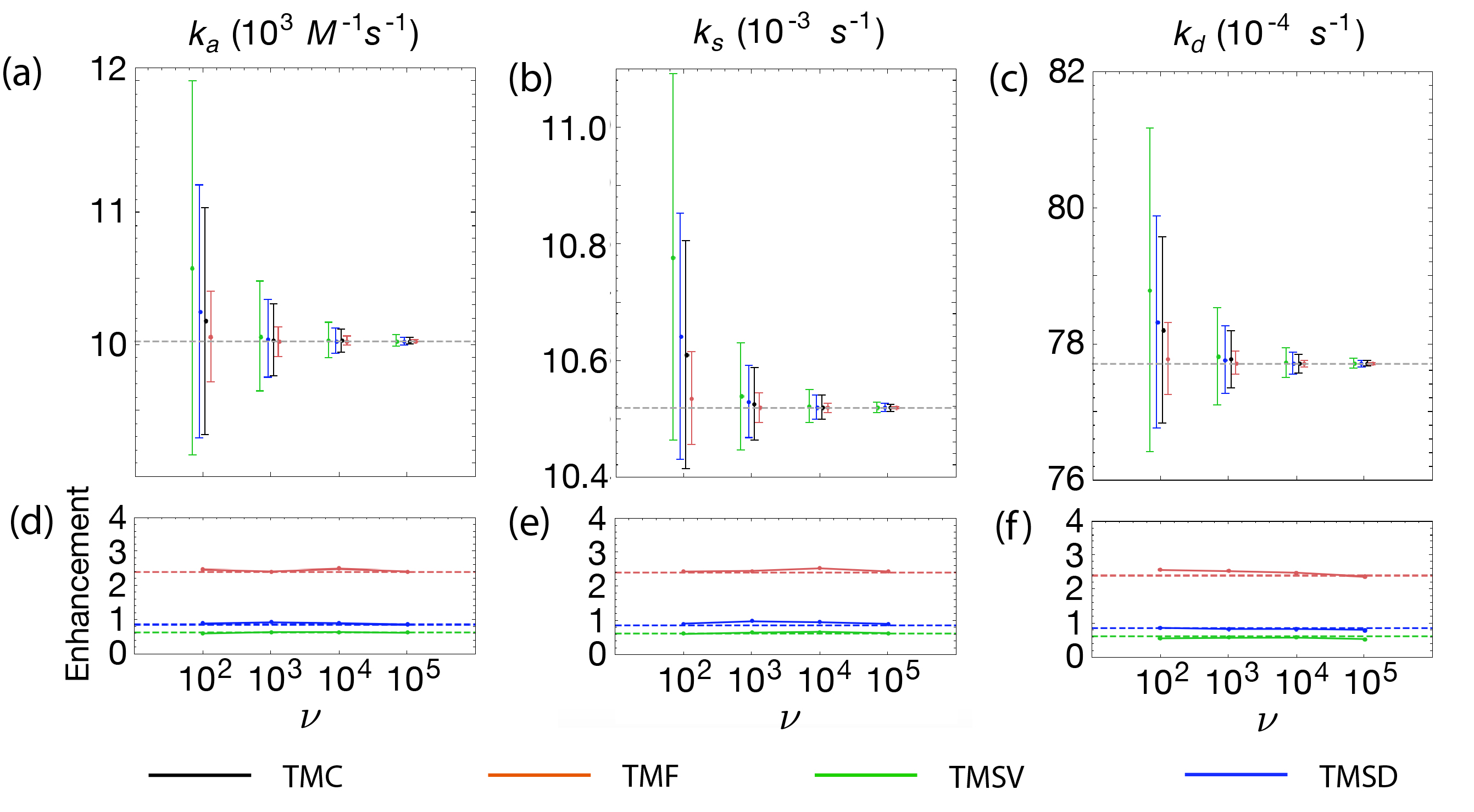}
\caption{Standard two-mode sensing using different quantum states (no loss, $\eta_a=\eta_b=1$). Panels (a), (b) and (c) show the estimation values and precisions for the kinetic parameters $k_a$, $k_s$ and $k_d$ as $\nu$ increases for $m=50$. For each value of $\nu$ the error bars represent the TMSV, TMSD, TMC and TMF states, going from left to right. Panels (d), (e) and (f) show the corresponding enhancement ratio for the different quantum states for $m=50$. From top to bottom the lines correspond to TMF, TMSD and TMSV, respectively. The dotted lines are a guide representing the enhancement expected from the ratio $R_M$ at the mid-point of the sensorgram for the respective state.}
\label{standardtwomodeKausaitem} 
\end{figure*}

In Fig.~\ref{standardtwomodeKausaite}(a)-(f) in the main text we showed the estimation value and precision for the kinetic parameters for $m=10$ sensorgrams in a set. In Fig.~\ref{standardtwomodeKausaitem}(a)-(c) we show the estimation value and precision for $m=50$ sensorgrams in a set. A similar behavior can be seen for all the states, with the TMF state providing the best estimation in the kinetic parameters for any $\nu$, followed by the TMC state, then the TMSD state and finally the TMSV state. The enhancement also behaves similarly to the case of $m=10$, as shown in Fig.~\ref{standardtwomodeKausaitem}(d)-(f).

\begin{figure}[t]
\centering
\includegraphics[width=16cm]{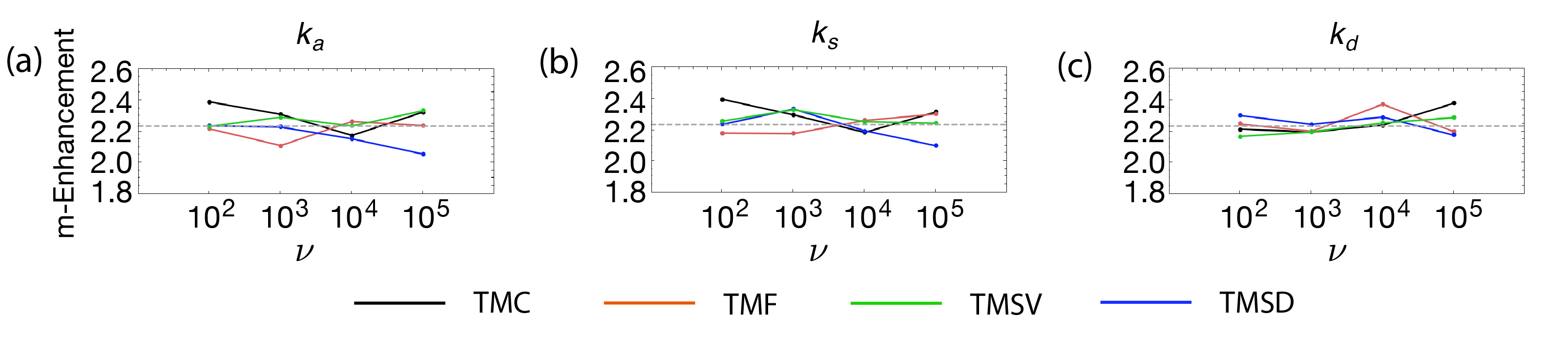}
\caption{The influence of $m$ in standard two-mode sensing using different quantum states (no loss, $\eta_a=\eta_b=1$). (a) $m$-enhancement ratio for $k_a$. From top to bottom (for first data point) the lines correspond to TMC, TMSD, TMSV and TMF, respectively. (b) $m$-enhancement ratio for $k_s$. From top to bottom (for first data point) the lines correspond to TMC, TMSV, TMSD and TMF, respectively. (c) $m$-enhancement ratio for $k_d$. From top to bottom (for first data point) the lines correspond to TMSD, TMF, TMC and TMSV, respectively. The dotted line corresponds to the expected ratio $\sqrt{50/10}=2.236$ at any point of the sensorgram.}
\label{standardtwomodeKausaitemcheck} 
\end{figure}

In Fig.~\ref{standardtwomodeKausaitemcheck} we show the ratio of the enhancement ratios for $m=50$ and $m=10$, {\it i.e.}, $R_{k,50}/R_{k,10}$, for $k_a$, $k_s$ and $k_d$ as $\nu$ increases. As mentioned in the main text, we call this the `$m$-enhancement' ratio. The dotted lines are a guide that represent the enhancement $R_{M,m}$ expected from the ratio of $\Delta M$ at any point of the sensorgram. The ratio of the enhancement ratios is expected to be $\sqrt{50/10}=2.236$. This is due to the $1/\sqrt{m}$ dependence of the estimation precision, $\Delta k$, for a fixed $\nu$. The $m$-enhancement ratio for the TMC is now also shown as it is expected to be roughly $\sqrt{50/10}$ when going from $m=10$ to $m=50$ sensorgrams in a set. 

\section{Enhancement in the precision around the sensorgram mid-point \label{sec:midpointenhance}}

\subsection{Standard two-mode scenario}

\begin{figure*}[t]
\centering
\includegraphics[width=18cm]{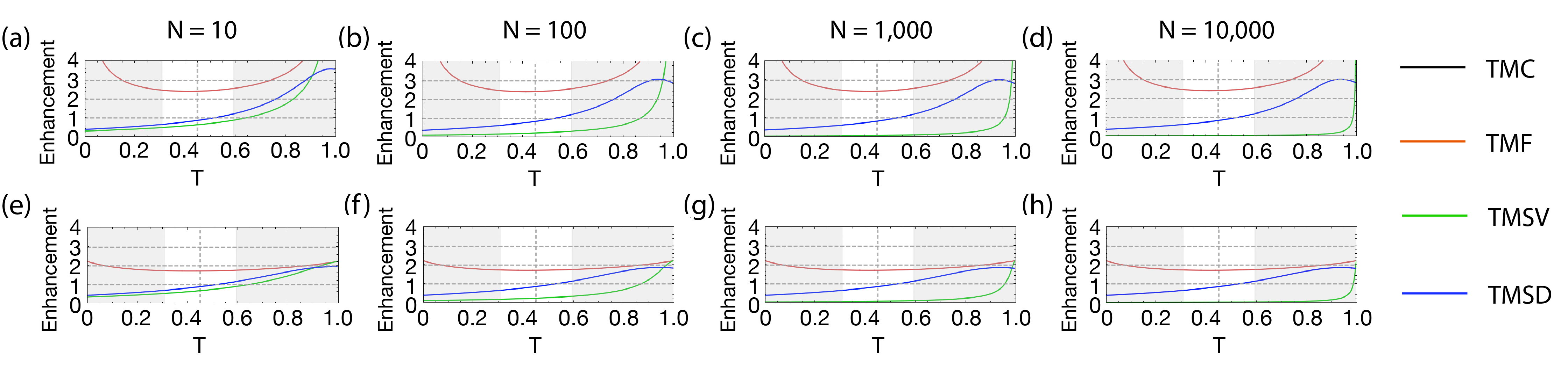}
\caption{Enhancement $R_M$ about the mid-point $T$ value for the standard two-mode scenario. (a)-(d) No loss ($\eta_a=\eta_b=1$). (e)-(h) Loss ($\eta_a=\eta_b=0.8$). In all panels the parameters are $\nu=100$ and $m=10$. From top to bottom the lines correspond to TMF, TMSD and TMSV, respectively. The white region corresponds to the maximum variation of $T$ for the sensorgram.}
\label{midpointstandardKausaite} 
\end{figure*}

Here we show how the enhancement $R_M$ changes about the mid-point of the sensorgram for each of the states in the standard two-mode scenario. In Fig.~\ref{midpointstandardKausaite}(a)-(d) we show the case of no loss ($\eta_a=\eta_b=1$) as the photon number $N$ increases. The general trend is that $T$ values below (above) the mid-point give a lower (higher) enhancement. The overall effect on the estimation precision of kinetic parameter $k$ is an averaging of the enhancement, leading to $R_k$. In Fig.~\ref{midpointstandardKausaite}(e)-(h) we show the case of loss ($\eta_a=\eta_b=0.8$) as $N$ increases. The general trend for $T$ follows the lossless case, although the enhancements are reduced -- most notably for the TMF state. In the case of no loss and loss there is no dependence on the photon number except for the TMSV state. The plots are obtained using Eqs.~\eqref{DeltaTMC}, \eqref{DeltaTMF}, \eqref{DeltaTMSV} and \eqref{DeltaTMSD} from Appendix \ref{sec:noise}.

\subsection{Optimized two-mode scenario}

\begin{figure*}[t]
\centering
\includegraphics[width=18cm]{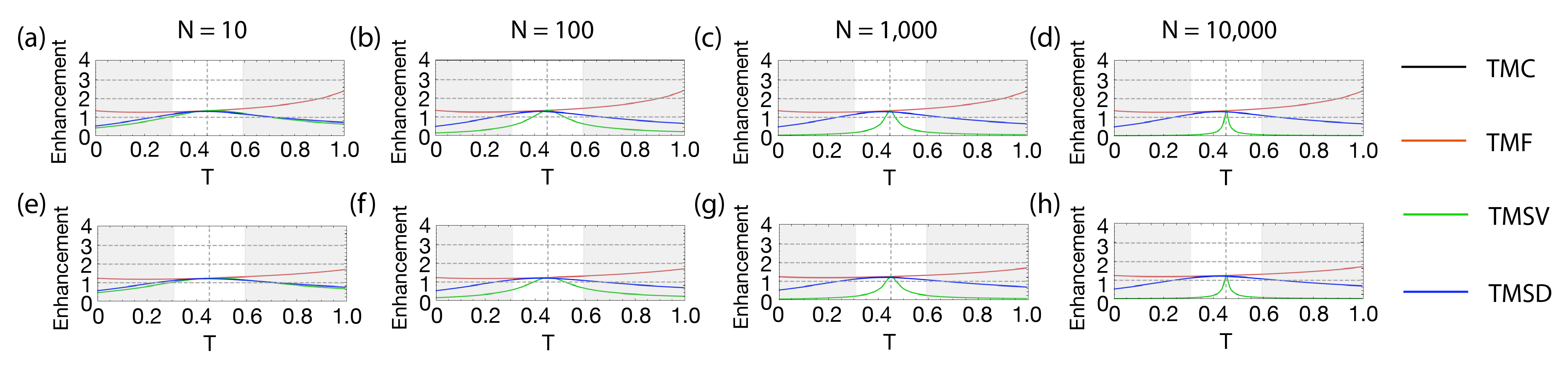}
\caption{Enhancement $R_M$ about the mid-point $T$ value for the optimized two-mode scenario. (a)-(d) No loss ($\eta_a=1$ and $\eta_b=\eta_aT_{mid}$). (e)-(h) Loss ($\eta_a=0.8$ and $\eta_b=0.8 T_{mid}$). In all panels the parameters are $\nu=100$ and $m=10$. From top to bottom the lines correspond to TMF, TMSD and TMSV, respectively. The white region corresponds to the maximum variation of $T$ for the sensorgram.}
\label{midpointoptimizedKausaite} 
\end{figure*}

Here we show how the enhancement $R_M$ changes about the mid-point of the sensorgram for each of the states in the optimized two-mode scenario. In Fig.~\ref{midpointoptimizedKausaite}(a)-(d) we show the case of no loss ($\eta_a=1$ and $\eta_b=T_{mid}$) as the photon number $N$ increases. Unlike the standard two-mode scenario, the enhancements for $T$ values around the mid-point are roughly constant. In Fig.~\ref{midpointoptimizedKausaite}(e)-(h) we show the case of loss ($\eta_a=0.8$ and $\eta_b=0.8T_{mid}$) as $N$ increases. The general trend for $T$ follows the lossless case, although the enhancements are slightly reduced. In the case of no loss and loss there is no dependence on the photon number except for the TMSV state. Unlike the standard two-mode scenario all states are affected similarly by the loss. The impact of the sharp downwards trend of the enhancement on either side of the mid-point for the TMSV state is responsible for overall TMSV state enhancement drop from that expected at the mid-point in the main text, as seen in Fig.~\ref{optimizedtwomodeKausaite}(j), (k) and (l).
\newpage
\phantom{.}

\end{document}